\documentclass{amsart}
\usepackage{epsfig}
\usepackage{amsmath,amssymb,amsfonts,amsthm}
\usepackage{mathrsfs}
\usepackage{youngtab}
\usepackage[hang,nooneline]{subfigure}

\numberwithin{equation}{section}
\newtheorem{theorem}{Theorem}[section]

\newtheorem{corollary}[theorem]{Corollary}
\newtheorem{proposition}[theorem]{Proposition}

\allowdisplaybreaks \numberwithin{equation}{section}

\newcommand{\vek}[1]{\boldsymbol{#1}}

\newcommand{\M}{\phantom{-}}
\newcommand{\weglassen}[1]{}

\renewcommand{\imath}{\mathrm{i}}
\renewcommand{\d}{\mathrm{d}}

\newcommand{\tla}{\tilde{\Lambda}}
\newcommand{\tl}{\tilde{L}}
\newcommand{\tr}{\tilde{r}}
\newcommand{\tq}{\tilde{q}}

\begin{document}

\title[Hyperelliptic functions]{Inversion of hyperelliptic integrals of arbitrary genus with application to particle motion in General Relativity}

\author{V.Z. Enolski}
\address{ZARM, University Bremen, Center of Applied Space Technology and Microgravity, Am Fallturm, 28359 Bremen, Germany, and Institute for Advanced Study, 27733 Delmenhorst, Germany, and Institute of Magnetism, National Academy of Sciences of Ukraine, Kiev} \email{vze@ma.hw.ac.uk}

\author{E. Hackmann}
\address{ZARM, University Bremen, Center of Applied Space Technology and Microgravity, Am Fallturm, 28359 Bremen, Germany }\email{eva.hackmann@zarm.uni-bremen.de}

\author{V. Kagramanova}
\address{Institute for Physics, University Oldenburg, 26111 Oldenburg, Germany}
\email{kavageo@theorie.physik.uni-oldenburg.de}

\author{J. Kunz}
\address{Institute for Physics, University Oldenburg, 26111 Oldenburg, Germany}
\email{kunz@theorie.physik.uni-oldenburg.de}

\author{C. L\"ammerzahl}
\address{ZARM, University Bremen, Center of Applied Space Technology and Microgravity, Am Fallturm, 28359 Bremen, Germany, and Institute for Physics, University Oldenburg, 26111 Oldenburg, Germany}
\email{laemmerzahl@zarm.uni-bremen.de}

\begin{abstract}
The description of many dynamical problems like the particle motion in higher dimensional spherically and axially symmetric space-times is reduced to the inversion of a holomorphic hyperelliptic integral. The result of the inversion is defined only locally, and is done using the algebro-geometric techniques of the standard Jacobi inversion problem and the foregoing restriction to the $\theta$--divisor. For a representation of the hyperelliptic functions the Klein--Weierstra{\ss} multivariable sigma function is introduced. It is shown that all parameters needed for the calculations like period matrices and Abelian images of branch points can be expressed in terms of the periods of holomorphic differentials and theta-constants. The cases of genus two and three are considered in detail. The method is exemplified by particle motion associated with a  genus three hyperelliptic curve.
\end{abstract}

\maketitle

{\small\tableofcontents}


\section{Introduction}

In a wide range of dynamical problems of classical systems one faces the problem of the inversion of integrals of the type \cite{Goldstein02}
\begin{equation}
t - t_0 = \int_{x_0}^x \frac{y^k}{\sqrt{\mathcal{P}_n}} dy
\end{equation}
where $\mathcal{P}_n$ is a polynomial of order $n$. Such so--called hyperelliptic integrals of genus $g$, $n = 2g+1$, are found, for example, for the particle motion in higher dimensional axially symmetric space--times. The solution of the inversion is the function $x = x(t)$. If the genus is $g=1$ then the result of the inversion is an elliptic function, which is a doubly periodic function of one complex variable. For higher genera $g > 1$ such an inversion becomes impossible because, as already recognized by Jacobi  (see e.g. \cite{marku92}), $2g$--periodic functions of one variable do not exist. However, Jacobi was able to resolve this contradiction by formulating his celebrated {\em Jacobi inversion problem} that involves $g$ hyperelliptic integrals.  The problem was solved in terms of so--called hyperelliptic functions which are indeed $2g$--periodic functions for $g>1$ which depend on $g$ variables while the periods (also called moduli) are $g \times g$--matrices. The domain of these hyperelliptic functions - the {\em Jacobi variety} - is thus the $g$--dimensional complex space $\mathbb{C}^g$ factorized by the period lattice.

The Jacobi inversion problem stimulated the development of algebraic geometry and in particular led to the discovery of solutions of many classical mechanical systems like Neumann's geodesic on an ellipsoid, the spinning top of Kowalewskaja, Kirchhoff's motion of a rigid body in a fluid, and others that were integrated using Jacobi's procedure. This special type of integrability that might be called {\it algebro--geometric integrability} has been receiving much attention owing to the discovery of the vast class of partial differential equations of  Korteveg--de Vries type that admits this type of integrability.

In this paper we consider the problem of the inversion of one hyperelliptic integral on the basis of the well developed algebro-geometric technique for the standard Jacobi inversion problem. The results of such inversions can be obtained by a restriction of hyperelliptic functions to special subsets of the Jacobi variety -- the {\em $\theta$--divisor}, that is given as a solution of the equation including the Riemann {\em $\theta$--function}. Such restrictions of hyperelliptic functions can be defined only locally; alternatively they can be realized on an infinitely sheeted Riemann surface \cite{fg07}. These functions inherit a number of properties of standard elliptic functions, like the addition formulae of the Frobenius--Stickelberger type \cite{on02}.

Our development starts with the standard Jacobi inversion problem that involves $g$ hyperelliptic integrals with variable bounds called {\em divisor} and describes a dynamic system with $g$ degrees of freedom. Then we are fixing $1<m<g$ points of the divisor making it {\em special} or a {\em divisor with deficiency}. In the context of this paper that means that we are considering the case when the genus of the underlying algebraic curve exceeds the number of degrees of freedom of the system. Although different values of $m$ appear in various problems we are concentrating here on the case of maximal deficiency $m=g-1$, i.e., on the inversion of one hyperelliptic integral.

The Jacobi inversion problem for a divisor with deficiency has a long history that includes Baker's consideration \cite{ba907}, Grant's \cite{gr90} and Jorgenson's \cite{jo92} treatment of the genus two case, \^Onishi's consideration \cite{on98} of the genus three case, description of certain dynamic systems with separable variables \cite{eekl93}, the treatment of the weak Kowalevski-Painlev\'e property \cite{abendfed00}, the integration of Somos sequences \cite{mat03}, \cite{beh05}, \cite{Hone07}, reductions of Benney hierarchies \cite{bg04},\cite{bg06} and others.
Here we will consider the problem of inversion in a systematic way within the Klein--Weierstra{\ss} realization of the theory of Abelian functions that is documented in the book of Baker \cite{ba97} (see also the review \cite{BEL97} and the more recent developments in Buchstaber and Leykin \cite{BL05}, also Nakayashiki \cite{nakaya08}, Matsutani and Previato \cite{matprev10}). The inversion formulae discussed here result in the restriction of the solution of the standard Jacobi inversion problem written in terms of the Kleinian $\wp$--functions to the corresponding stratum of the $\theta$--divisor. In this context there appears the problem of a suitable parametrization of the $\theta$--divisor in the case of higher genera. We solve this problem on the basis of Newton's method for the approximation of multivariable functions. We note that in this paper we are discussing the inversion of holomorphic integrals only; similar considerations can be undertaken for meromorphic integrals and integrals of the third kind, see e.g. \cite{EPR03}.

In order to elaborate an effective calculation procedure for the inversion of one holomorphic hyperelliptic integral we are solving a problem of general interest, that is, the calculation of the period matrix of meromorphic differentials that is sometimes called {\em second period matrix}. Existing Maple codes contain the evaluation of the Riemann period matrix by the given curve, i.e., period matrices of holomorphic differentials, only. However, the calculation of the periods of meromorphic differentials is unavoidable in certain problems and in particular in our inversion problem. Therefore we show that it is possible to express this second period matrix in terms of the first period matrix and theta-constants. In this way we reduce the number of complete integrals, which are necessary to calculate, to the Riemann period matrix given by Maple codes. Another result yields the characteristics of the Abelian images of branch points by the given holomorphic period matrices and, thus, that the homology basis can be reconstructed by these data. That permits to find the vector of Riemann constants and to carry out the whole calculation without referring   to the homology basis. While $\theta$--functional calculations are usually considered as technically complicated, here we describe an easily algorithmized scheme that turn such calculations into routine procedures at least in the hyperelliptic case.

Our paper is organized as follows. In Section~\ref{sec:particlemotion} we shortly describe the problem of particle motion in higher dimensional spherically and axially symmetric space-times. This problem will serve as a laboratory for the approbation of the methods developed. In Section~\ref{secPre} we recall the known facts from the theory of hyperelliptic functions and develop a realization of these functions in terms of Klein--Weierstra{\ss} multivariable $\sigma$--function. Special attention is focused on the effective calculation of the moduli of the system. We show in particular how to express periods of meromorphic differentials in terms of the theta-constants and periods of holomorphic differentials. In Section~\ref{sec:inversionhei} we are considering the stratification of the $\theta$--divisor and show how to single out the stratum that is the image of the curve inside the Jacobian in terms of conditions on the $\sigma$--function. This stratum serves as the domain for the quasi-elliptic function. In Sections~\ref{sec:curvesgenus2} and~\ref{sec:curvesgenus3} we are considering the application of the method developed to the cases of genus two and genus three hyperelliptic curves. Finally in the last Section~\ref{sec:application} we are coming back to the initial physical problem and demonstrate how to compute the trajectories of test particles in higher dimensions by the method of restriction to the $\theta$--divisor. We explicitely calculate orbits in a $9$-dimensional Reissner--Nordstr\"om--de Sitter space--time, which is characterized by its mass, electric charge and the cosmological constant. The underlying polynomial is of degree $7$ which corresponds to a genus $3$ hyperelliptic curve. This is the generalization of the examples considered in~\cite{Hackmannetal08}, where orbits with underlying hyperelliptic curves of genus $2$ where calculated.

We believe that our method has much wider applications than the special physical problem considered here and that it can be used in other problems that needs the inversion of a hyperelliptic integral.
The same approach works for the inversion of meromorphic integrals, that we will consider elsewhere. Some of our results can be used for the Jacobi inversion problem on the strata with smaller deficiency.

\section{Particle motion in General Relativity}\label{sec:particlemotion}

Ordinary differential equations of the form
\begin{equation}
\frac{dx}{dt} = f(x, \sqrt{\mathcal{P}_4(x)})\,,
\end{equation}
where $f$ is a rational function of $x$ and $\mathcal{P}_4(x)$ is a polynomial of order four are solved by elliptic integrals introduced by Jacobi and Weierstra{\ss}\footnote{Here and below we will closely follow the standard notations of the theory of elliptic functions fixed in
 \cite{be55}.}. The corresponding equations of that form with a fourth order polynomial $P_4(x)$ can be reduced to one with a third order polynomial. As an example we mention the motion of a point particle or light ray given by the geodesic equation
\begin{equation}
\frac{d^2 x^\mu}{ds^2} + \left\{\begin{smallmatrix} \mu \\ \rho\sigma \end{smallmatrix}\right\} \frac{d x^\rho}{ds} \frac{d x^\sigma}{ds} = 0
\end{equation}
with the Christoffel symbol
\begin{equation}
\left\{\begin{smallmatrix} \mu \\ \rho\sigma \end{smallmatrix}\right\} := \frac{1}{2} g^{\mu\nu} \left(\partial_\rho g_{\sigma\nu} + \partial_\sigma g_{\rho\nu} - \partial_\nu g_{\rho\sigma}\right) \, ,
\end{equation}
where $g_{\mu\nu}$ is the space--time metric. $ds$ is the proper time defined by $ds^2 = g_{\mu\nu} dx^\mu dx^\nu$. For a light--like particle (photon) the parameter $s$ is replaced by some affine parameter.

For a Schwarzschild metric
\begin{equation}
ds^2 = g_{tt} dt^2 - g_{rr} dr^2 - r^2 \left(d\vartheta^2 + \sin^2\vartheta d\varphi^2\right) \,
\end{equation}
with
\begin{equation}
g_{tt} = \frac{1}{g_{rr}} = 1 - \frac{2 M}{r} \, ,
\end{equation}
where $M$ is the mass of the gravitating body (we choose units so that the Newtonian gravitational constant as well as the velocity of light are unity, $G = c = 1$), this geodesic equation with a substitution $x=f(r)$ yields an equation describing the dependence of the radial coordinate $r$ on the azimuthal angle $\varphi$
\begin{equation}
\left(\frac{dx}{d\varphi}\right)^2 = 4 x^3 - g_2 x - g_3 \, ,
\end{equation}
where the Weierstra{\ss} invariants $g_2$ and $g_3$ depend on $M$ and the energy and angular momentum of the particle. The complete set of solutions in terms of the Weierstra{\ss} $\wp$--function have been given and extensively discussed by Hagihara \cite{Hagihara31}. The corresponding periods of the $\wp$--functions are directly related to observable effects like the perihelion shift and the deflection (scattering) angle of massive bodies and of light.

Geodesic equations of neutral test particles or photons in Taub-NUT space-times~\cite{KKHL10} and of charged test particles in Reissner-Nordstr\"om space-times~\cite{GK10} are also solved in terms of elliptic Weierstra{\ss} functions. In the case of a Schwarzschild--(anti-)de Sitter or Kerr--(anti-)de Sitter metric we encounter similar equations but with a polynomial of fifth and sixth order (where the sixth order polynomial can be reduced to a fifth order one)
\begin{equation}
\left(x^i\frac{dx}{dt}\right)^2 = \mathcal{P}_5(x) \ , \, i=0,1 \ , \quad \text{or} \ , \quad \left(\frac{dx}{dt}\right)^2 = (x-c)^2\mathcal{P}_5(x) \ ,
\end{equation}
where $c$ is a constant.
The corresponding equations have been solved explicitely in \cite{HackmannLaemmerzahl08,HackmannLaemmerzahl08a,Hackmannetal09,Hackmannetal010}.

These examples can be generalized further to include polynomials of even higher orders as outlined in the following.

\subsection{Geodesic equations in higher dimensional spherically symmetric space--times}

The metric of a spherically symmetric Reissner--Nordstr\"om--(anti-)de Sitter space--times in $d$ dimensions is given by
\begin{equation}
ds^2 = g_{tt} dt^2 - g_{rr} dr^2 - r^2 d\Omega^2_{d-2} \, ,
\end{equation}
with
\begin{equation}
g_{tt} = \frac{1}{g_{rr}} = 1-\left(\frac{r_{\rm S}}{r}\right)^{d-3}-\frac{2\Lambda r^2}{(d-1)(d-2)} + \left(\frac{q}{r}\right)^{2(d-3)} \, ,
\end{equation}
where $\Lambda$ is the cosmological constant, $q$ the charge of the gravitating mass $M$, $r_{\rm S}=2 M$, and $d\Omega^2_{d-2}$ is the surface element of the $d-2$--dimensional unit sphere. The geodesic equation then leads to
\begin{align}
\left(\frac{dr}{d\varphi}\right)^2 & = \frac{r^4}{L^2} \frac{1}{g_{rr} g_{tt}} \left(E^2 - g_{tt} \left(\delta + \frac{L^2}{r^2}\right)\right)\\
& = \frac{r^4}{L^2} \left(E^2 - \left(1-\left(\frac{r_{\rm S}}{r}\right)^{d-3}-\frac{2\Lambda r^2}{(d-1)(d-2)} + \left(\frac{q}{r}\right)^{2(d-3)} \right) \left(\delta + \frac{L^2}{r^2}\right)\right) \label{beglkusy} \ ,
\end{align}
where $E$ and $L$ are two conserved quantities: the dimensionless energy $E$, and the angular momentum $L$ with the dimension of length (both are normalized to the mass of a test particle)
\begin{equation}
E= g_{tt} \frac{dt}{d\lambda} \ , \qquad L = r^2\frac{d\varphi}{d\lambda} \ ,
\end{equation}
where $\lambda$ is an affine parameter along the geodesic. $\delta=1$ for massive test particles and $\delta=0$ for massless particles.
A substitution $x = f(r)$ gives equations of the form
\begin{equation}
\left(x^{i}\frac{dx}{d\varphi}\right)^2 = \mathcal{P}_n(x)
\end{equation}
for some $0 \leq i < g$ where $\mathcal{P}_n$ denotes a polynomial of order $n$ and $g=\left[ \frac{n+1}{2} \right]$ is the genus of a curve $w^2=\mathcal{P}_n(x)$. In some cases through appropriate substitutions the order of the polynomial can be reduced \cite{Hackmannetal08}. However, in general we have $n \geq 7$, as in the example in Section~\ref{sec:application}.

\subsection{The effective one--body problem}

Another example of spherically symmetric problems is the relativistic effective one--body problem in four dimensions. While in Newtonian gravity the two--body problem can be exactly reduced to an one--body problem this is not possible in Einstein's General Relativity. The relativistic two--body problem can be reduced to a one--body problem only in terms of a series expansion. In this framework the relative coordinate between two bodies with masses $M_1$ and $M_2$ formally fulfills a geodesic equation in a space--time with the effective metric
\begin{equation}
ds^2 = - g_{tt}(r, \nu) dt^2 + g_{rr}(r, \nu) dr^2 + r^2 (d\vartheta^2 + \sin^2\vartheta d\varphi^2) \, ,
\end{equation}
where $u = 2 (M_1 + M_2)/r$, $\nu = M_1 M_2/(M_1 + M_2)^2$ and
\begin{align}
\begin{split}
g_{tt}(r,\nu) & = 1 - 2 u + 2 \nu u^3 + \nu a_4 u^4 + \mathscr{O}(u^5) \\
\left(g_{tt}(r, \nu) g_{rr}(r, \nu)\right)^{-1} & = 1 + 6 \nu u^2 + 2 (26 - 3 \nu) \nu u^3 + \mathscr{O}(u^4) \, .
\end{split}
\end{align}
The corresponding effective one--body equation of motion is then given by \cite{BuonannoDamour99,DamourJaranowskiSchaefer00a}
\begin{equation}
\left(\frac{dr}{d\varphi}\right)^2 = \frac{r^4}{L^2} \frac{1}{g_{rr} g_{tt}} \left(E^2 - g_{tt} \left(1 + \frac{L^2}{r^2}\right)\right) \, ,
\end{equation}
where $E$ and $L$ are again the conserved energy and angular momentum. This is a  series expansion which can be expanded to arbitrary order. Although this is only a series expansion, analytic methods are helpful for the purpose to have a complete discussion of the possible types of orbits of a binary system.

\subsection{Geodesic equations in higher dimensional axially symmetric space--times}

As an example of $d$--dimensional axially symmetric space--times we consider the simplest one, namely the Myers--Perry~\cite{MyersPerry1986} space--times with only one rotation parameter $a$ given by \cite{KodamaKonoplya2010,Emparan2008}
\begin{align}
ds^2 = & \frac{1}{\rho}\left(\frac{2M}{r^{n-1}} - \rho^2 \right) dt^2 - \frac{4aM\sin^2\theta}{\rho^2 r^{n-1}} dt d\varphi + \frac{\sin^2\theta}{\rho^2} \left((r^2 + a^2) \rho^2 + \frac{2a^2M}{r^{n-1}}\sin^2\theta \right) d\varphi^2 \nonumber \\
& + \frac{\rho^2}{\Delta} dr^2 + \rho^2 d\theta^2 + r^2 \cos^2\theta d\Omega^2_n \ ,
\end{align}
where $\rho^2=r^2+a^2\cos^2\theta$, $\Delta=(r^2+a^2)-\frac{2M}{r^{n-1}}$, and $n=d-4$. Here, $M$ is the mass of the black hole and the surface element of the unit $n$-sphere is $d\Omega^2_n = \sum^n_{i=1} \prod^{i-1}_{k=1} \sin^2\psi_k d\psi^2_i$.

Owing to the conservation laws related to the symmetries of the underlying space--time there are a conserved energy $E$, an angular momentum $L_\varphi$ and $n$ further constants $\Psi^2_i$, $i=1,\ldots,n$. As a consequence, it is possible to separate the Hamilton--Jacobi equation~\cite{vasupa2005}. The resulting equations of motion are then given by
\begin{eqnarray}
\rho^2 \frac{dr}{d\lambda} & = & \sqrt{R} \label{hj-r2} \\
\rho^2 \frac{d\theta }{d\lambda } & = & \sqrt{\Theta} \,  \label{hj-theta2} \\
\phantom{\rho^2} \frac{d\varphi}{d\lambda} & =  & \frac{a}{\Delta} \left[(r^2+a^2)E - aL_\varphi\right] \frac{1}{\sqrt{R}}\frac{dr}{d\lambda} + \frac{1}{\sin^2\theta}\left[L_\varphi - a\sin^2\theta E \right] \frac{1}{\sqrt{\Theta}}\frac{d\theta}{d\lambda}
 \label{hj-phi2} \\
\phantom{\rho^2} \frac{d t}{d\lambda} & = & \frac{r^2+a^2}{\Delta} \left[(r^2+a^2)E - aL_\varphi\right] \frac{1}{\sqrt{R}}\frac{dr}{d\lambda} +  a \left[L_\varphi - a\sin^2\theta E \right] \frac{1}{\sqrt{\Theta}}\frac{d\theta}{d\lambda} \, .  \label{hj-t2} \\
\phantom{\rho^2} \frac{d \psi_{i}}{d\lambda} & = & \frac{\sqrt{A_{i}}}{\prod^{i-1}_{k=1}\sin^2\psi_{k}} \frac{1}{r^2\cos^2\theta} \ , i=1,..,n \, \label{hj-psin2} \ ,
\end{eqnarray}
with $A_i=\Psi^2_i - \frac{\Psi^2_{i+1}}{\sin^2\psi_i}$ and
\begin{align}
R(r) & = \left[(r^2+a^2)E - aL_\varphi\right]^2  - \Delta \left( K + \delta r^2 + \frac{a^2}{r^2} \Psi^2_1 \right) \,, \\
\Theta(\theta) & = K - \delta a^2\cos^2\theta - \frac{\Psi^2_1 }{\cos^2\theta}  - \frac{1}{\sin^2\theta} \left[L_\varphi - a\sin^2\theta E \right]^2 \, ,
\end{align}
where $K$ is a further constant, called the Carter constant, which emerges from the separation process. It is obvious that $\mathcal{P}(r)$ in $R(r)=\frac{\mathcal{P}_{d+1}(r)}{r^{d-3}}$ is a polynomial whose order increases with the dimension of the space--time. (In some cases the order of $\mathcal{P}(r)$ can be reduced by a substitution.)

In the following sections we will explain the theory and the details of how to analytically solve the geodesic equations in the above cases and for similar physical problems.

\section{Hyperelliptic functions}\label{secPre}

\begin{figure}
\begin{center}
\unitlength 0.7mm \linethickness{0.4pt}
\begin{picture}(150.00,80.00)
\put(-11.,33.){\line(1,0){12.}} \put(-11.,33.){\circle*{1}}
\put(1.,33.){\circle*{1}} \put(-10.,29.){\makebox(0,0)[cc]{$e_1$}}
\put(1.,29.){\makebox(0,0)[cc]{$e_2$}}
\put(-5.,33.){\oval(20,30.)}
\put(-12.,17.){\makebox(0,0)[cc]{$\mathfrak{ a}_1$}}
\put(-5.,48.){\vector(1,0){.7}}
\put(12.,33.){\line(1,0){9.}} \put(12.,33.){\circle*{1}}
\put(21.,33.){\circle*{1}} \put(13.,29.){\makebox(0,0)[cc]{$e_3$}}
\put(22.,29.){\makebox(0,0)[cc]{$e_4$}}
\put(17.,33.){\oval(18.,26.)}
\put(10.,19.){\makebox(0,0)[cc]{$\mathfrak{ a}_2$}}
\put(16.,46.){\vector(1,0){.7}}
\put(35.,33.){\circle*{1}} \put(40.,33.){\circle*{1}}
\put(45.,33.){\circle*{1}}

\put(60.,33.){\line(1,0){9.}} \put(60.,33.){\circle*{1}}
\put(69.,33.){\circle*{1}}
\put(60.,29.){\makebox(0,0)[cc]{$e_{2g-1}$}}
\put(72.,29.){\makebox(0,0)[cc]{$e_{2g}$}}
\put(65.,33.){\oval(30,15.0)}
\put(59.,20.){\makebox(0,0)[cc]{$\mathfrak{ a}_g$}}
\put(62.,40.2){\vector(1,0){.7}}
\put(114.,33.00){\line(1,0){33.}} \put(114.,33.){\circle*{1}}
\put(147.,33.){\circle*{1}}
\put(115.,29.){\makebox(0,0)[cc]{$e_{2g+1}$}}
\put(146.,29.){\makebox(0,0)[cc]{$e_{2g+2}=\infty$}}
\put(18.,72.){\makebox(0,0)[cc]{$\mathfrak{ b}_1$}}
\put(25.,78.){\vector(4,1){0.2}}
\bezier{484}(-4.,33.00)(0.,76.)(25.,78.)
\bezier{816}(25.00,78.)(75.00,82.00)(143.00,33.00)
\put(33.,64.){\makebox(0,0)[cc]{$\mathfrak{ b}_2$}}
\put(38.00,70.){\vector(4,1){0.2}}
\bezier{384}(17.,33.00)(17.,68.)(38.00,70.)
\bezier{516}(38.00,70.)(80.00,74.00)(137.00,33.00)
\put(72.,48.){\makebox(0,0)[cc]{$\mathfrak{ b}_g$}}
\put(72.,54.){\vector(3,1){0.2}}
\bezier{226}(64.00,33.)(64.00,52.00)(72.,54.)
\bezier{324}(72.,54.)(88.,58.00)(125.00,33.00)
\end{picture}
\end{center}
\caption{A homology basis on a Riemann surface of the hyperelliptic curve of genus $g$ with real branch points $e_1,\ldots,e_{2g+2}=\infty$ (upper sheet).  The cuts are drawn from $e_{2i-1}$ to $e_{2i}$ for $i=1,\dots,g+1$.  The $\mathfrak b$-cycles are completed on the lower sheet (the picture on the lower
sheet is just flipped horizontally).} \label{figure-1}
\end{figure}
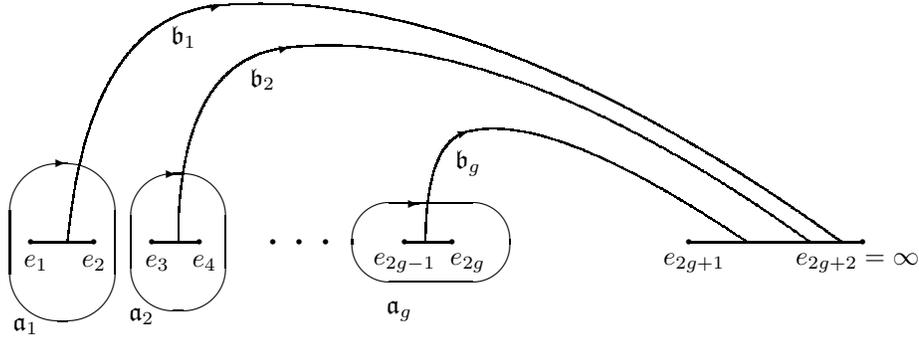

The solutions of the differential equations given in the foregoing section can be considered as points on a hyperelliptic curve $X_g$ of genus $g$ given by the equation
\begin{equation}\label{curve}
w^2= \mathcal{P}_{2g+1}(z) = \sum_{i=0}^{2g+1} \lambda_{i}z^{i} = 4\prod_{k=1}^{2g+1} (z-e_{k}),
\end{equation}
and realized as a two sheeted covering over the Riemann sphere branched in the points $(e_k,0)$, $k \in {\mathcal G} = \{1,\ldots,2g+1\}$, with $e_j\neq e_k$ for $j\neq k$, and at infinity, $e_{2g+2}=\infty$. Notice that we do not require the $e_k$ to be real. However, when they are real, we find it convenient to order them according to $e_1 < e_2 < \ldots < e_{2g+1}$, i.\,e., in the opposite way as compared to the Weierstra{\ss} ordering, see Fig.~\ref{figure-1}. Denote $P=(z,w)$ a coordinate of the curve. The factor 4 in (\ref{curve}) is introduced to preserve resemblance with the Weierstra{\ss} cubic for $g=1$,
\begin{equation}
w^2=4z^3-g_2z-g_3\equiv 4(z-e_1)(z-e_2)(z-e_3) \label{wcubic}.
\end{equation}

As shown in Fig.~\ref{figure-1} we equip the hyperelliptic curve $X_g$ with a canonical homology basis
\begin{align}
({\mathfrak a}_1,\ldots,{\mathfrak a}_g; {\mathfrak b}_1,\ldots, {\mathfrak b}_g), \quad\quad \mathfrak{a_i} \circ \mathfrak{b_j}=-\mathfrak{b_i} \circ \mathfrak{a_j}=\delta_{ij},\; \mathfrak{a_i}\circ \mathfrak{a_j}= \mathfrak{b_i}\circ \mathfrak{b_j}=0 \, ,
\end{align}
where $\delta_{ij}$ is the Kronecker symbol.

\subsection{Canonical differentials}

We choose canonical holomorphic differentials (of the first kind) $\d \vek{u}^t=(\d u_1,\ldots,\d u_g)$ and associated meromorphic differentials (of the second kind) $\d \vek{r}^t=(\d r_1,\ldots,\d r_g)$ in such a way that their $g \times g$ period matrices
\begin{align} \label{periods}\begin{split}
 2\omega  &= \Bigl(\M\oint_{{\mathfrak a}_k} \d u_i\Bigr)_{i,k=1,\ldots,g}, \qquad
 2\omega' = \Bigl(\M\oint_{{\mathfrak b}_k} \d u_i\Bigr)_{i,k=1,\ldots,g} \\
 2\eta    &= \Bigl(-\oint_{{\mathfrak a}_k}\d r_i\Bigr)_{i,k=1,\ldots,g} ,\qquad
 2\eta'   = \Bigl(-\oint_{{\mathfrak b}_k}\d r_i\Bigr)_{i,k=1,\ldots,g}\end{split}
\end{align}
satisfy the generalized Legendre relation
\begin{equation}
MJM^T=-\frac{\imath\pi}{2} J
 \label{Legendre}
\end{equation}
with
\begin{equation} \label{defM}
M = \begin{pmatrix}
\omega & \omega' \\ \eta & \eta' \end{pmatrix} \, , \qquad
J = \begin{pmatrix} 0 & -1_g \\ 1_g  & 0 \end{pmatrix} \, ,
\end{equation}
with $0_g$ and $1_g$ as the zero and unit $g \times g$--matrices. Such a basis of differentials can be realized as follows (see~Baker (1897), p.~195)
\begin{align}
\d \vek{u}(z,w) &= \frac{\boldsymbol{\mathcal U}(z)\d z}{w}, &
    {\mathcal U}_i( z) &= z^{i-1}, \qquad i=1\ldots,g,
  \label{uui}  \\
\d \vek{r}(z,w) &= \frac{\boldsymbol{\mathcal R}(z) \d z}{4w},&
    {\mathcal R}_i(z) &= \sum_{k=i}^{2g+1-i}(k+1-i)\lambda_{k+1+i}z^k, \qquad i=1\ldots,g \, ,
  \label{rrj}
\end{align}
where the coefficients $\lambda_i$ are given by (\ref{curve}).

We denote by $\mathrm{Jac}(X_g)$ the Jacobian of the curve $X_g$, i.e., the factor $\mathbb{C}^g/ \Gamma$, where $\Gamma=2 \omega \oplus 2 \omega'$ is the lattice generated by the periods of the canonical holomorphic differentials. Any point $\boldsymbol{u} \in \mathrm{Jac}(X_g)$ can be presented in the form
\begin{equation}
\boldsymbol{u}=2\omega \boldsymbol{\varepsilon}+2\omega' \boldsymbol{\varepsilon}' \, ,
\end{equation}
where $ \boldsymbol{\varepsilon},\boldsymbol{\varepsilon}'\in \mathbb{R}^g $. 
The vectors $\varepsilon$ and $\varepsilon^\prime$ combine to a $2 \times g$ matrix and form the characteristic of the point $\boldsymbol{u}$,
\begin{equation}
[\boldsymbol{u}] := \begin{pmatrix} \boldsymbol{\varepsilon'}^T \\
\boldsymbol{\varepsilon}^T \end{pmatrix} = \begin{pmatrix} \varepsilon_1' & \ldots & \varepsilon_g' \\
\varepsilon_1 & \ldots & \varepsilon_g \end{pmatrix} =: \varepsilon \, .
\end{equation}
If $\boldsymbol{u}$ is a half-period, then all entries of the characteristic $\varepsilon$ are equal to $\frac12$ or $0$.

Beside the canonical holomorphic differentials $\mathrm{d}\boldsymbol{u}$ we will also consider normalized holomorphic differentials defined by
\begin{equation}
 \label{eq:normdiff}
 \d \vek{v} = (2\omega)^{-1}\d \vek{u} \, .
\end{equation}
Their corresponding holomorphic periods are $1_g$ and $\tau$, where the Riemann period matrix $\tau := \omega^{-1} \omega'$ is in the Siegel upper half space $\mathfrak{S}_g$ of $g \times g$--matrices (or half space of degree $g$),
\begin{equation}
\mathfrak{S}_g = \left\{\tau \;\; g \times g \;\; \text{matrix} \big| \tau^T=\tau, \, \mathrm{Im}\,\tau \text{positive definite} \right\}\,.
\end{equation}
The corresponding Jacobian is introduced as
\begin{equation}
\widetilde{\mathrm{ Jac}}(X_g) := (2\omega)^{-1}\mathrm{ Jac} (X_g)=\mathbb{C}^g/ 1_g\oplus \tau \,.
\end{equation}
We will use both versions: the first one $(2\omega,2\omega')$ in the context of $\sigma$--functions, and the second one $(1_g,\tau)$ in the case of $\theta$--functions.

The Abel map $\boldsymbol{\mathfrak{A}}: (X_g)^n \rightarrow \mathbb{C}^g$ with the base point $P_0$ relates the set of points $(P_1,\ldots,P_n)$ (which are called the divisor $\mathcal{D}$) with a point in the Jacobian $\mathrm{ Jac}(X_g)$
\begin{equation}
\boldsymbol{\mathfrak{A}}(P_1,\ldots,P_n) := \sum_{k=1}^n  \int_{P_0}^{P_k} \d \boldsymbol{u} \,.
 \label{jip}
\end{equation}
The divisor $\mathcal{D}$ in (\ref{jip}) can be also denoted as $P_1+\ldots+P_n - n P_0$.

Analogously we define
\begin{equation}
\widetilde{\boldsymbol{\mathfrak{A}}}(P_1,\ldots,P_n)= \sum_{k=1}^n  \int_{P_0}^{P_k} \d \boldsymbol{v}= (2\omega)^{-1}\boldsymbol{\mathfrak{A}}(P_1,\ldots,P_n)\,.
 \label{jiptilde}
\end{equation}
In the context of our consideration we take $P_0$ as infinity, $P_0=(\infty,\infty)$.

\subsection{$\theta$--functions}\label{subs-theta}

The hyperelliptic $\theta$--function with characteristics $[\varepsilon]$ is a mapping $\theta:\ \widetilde{\mathrm{Jac}}(X_g) \times \mathfrak{S}_g \rightarrow \mathbb{C}$ defined through the Fourier series
\begin{equation}
\theta[\varepsilon] (\vek{ v}| \tau) := \sum_{\vek{m} \in \mathbb{Z}^g} e^{\pi \imath \left\{(\vek{m} + \vek{\varepsilon}')^t \tau (\vek{m} + \vek{\varepsilon}') + 2 (\vek{ v}+ \vek{\varepsilon})^t (\vek{m} + \vek{\varepsilon}')\right\}} \, .
\end{equation}
It possesses the periodicity property
\begin{equation}
\theta[\varepsilon] (\vek{v}+\vek{n} + \tau\vek{n}'|\tau) = e^{-2\imath\pi{\vek{n}'}^t(\vek{v} + \frac12\tau \vek{n}')} e^{2\imath\pi(\vek{n}^t\vek{\varepsilon}' - {\vek{n}'}^t\vek{\varepsilon})} \theta[\varepsilon](\vek{v}|\tau) \, . \label{thetachar2a}
\end{equation}
For vanishing characteristic we abbreviate $\theta(\vek{v}) := \theta[0] (\vek{v}| \tau)$.

In the following, the values $\varepsilon_k$, $\varepsilon_k'$ will either be $0$ or $\frac{1}{2}$. The equality (\ref{thetachar2a}) implies
\begin{equation}
\theta[\varepsilon]({-\vek v}|\tau)= \mathrm{e}^{-4\pi \imath
\vek{\varepsilon}^t\vek{\varepsilon}'} \theta[\varepsilon](\vek{v}|\tau)\label{-v},
\end{equation}
so that the function $\theta[\varepsilon](\vek{v}|\tau)$ with characteristics  $[\varepsilon]$ of only half-integers is even if $4\vek{\varepsilon}^t\vek{\varepsilon}'$ is an even integer, and odd otherwise. Correspondingly, $[\varepsilon]$ is called even or odd, and among the $4^g$ half-integer characteristics there are $\frac{1}{2} (4^g+2^g) $ even and $\frac{1}{2}(4^g-2^g) $ odd characteristics.

The non-vanishing values of the $\theta$--functions with half--integer characteristics and their derivatives are called $\theta$-constants and are denoted as
\begin{align*}
\theta[\varepsilon] & : =\theta[\varepsilon](\vek{0};\tau), & \quad  \theta_{i,j}[\varepsilon] & := \left.\frac{\partial^2}{\partial z_i\partial z_j}\theta[\varepsilon](\boldsymbol{z};\tau)\right|_{\boldsymbol{z}=0},\quad & \text{etc.} \qquad \text{for even $[\varepsilon]$};\\
\theta_{i}[\varepsilon] & :=\left.\frac{\partial}{\partial z_i}\theta[\varepsilon](\boldsymbol{z};\tau)\right|_{\boldsymbol{z}=0}, & \quad
\theta_{i,j,k}[\varepsilon] & := \left.\frac{\partial^3}{\partial z_i\partial z_j \partial z_k}\theta[\varepsilon](\boldsymbol{z};\tau)\right|_{\boldsymbol{z}=0},\quad
  & \text{etc.} \qquad \text{for odd $[\varepsilon]$} \,.
\end{align*}
Even characteristics $[\varepsilon]$ are called nonsingular if $\theta[\varepsilon]\neq 0$, and odd characteristics $[\varepsilon]$ called nonsingular if $\theta_i[\varepsilon]\neq 0$ at least for one index $i$.

We identify each branch point $e_j$ of the curve $X_g$ with a vector
\begin{equation}
{\boldsymbol{\mathfrak{A}}}_j := \int_{\infty}^{e_j} \d \vek{u} =: 2 \omega \boldsymbol{\varepsilon}_j + 2 \omega' \boldsymbol{\varepsilon}'_j \in \mathrm{Jac}(X_g),\quad j=1,\ldots, 2g+2 \, ,
\end{equation}
what defines the two vectors $\boldsymbol{\varepsilon}_j$ and $\boldsymbol{\varepsilon}'_j$. Evidently, $[{\boldsymbol{\mathfrak A}}_{2g+2}] = [0] = 0$.

In terms of the $2g+2$ characteristics $[{\mathfrak A}_i]$ all $4^g$ half integer characteristics $[\varepsilon]$  can be constructed as follows. There is a one-to-one correspondence between these $[\varepsilon]$ and partitions of the set $\bar{{\mathcal G}} = \{1, \ldots, 2g+2\}$ of indices of the branch points~(\cite{fa73}, p.~13, \cite{ba97} p.~271). The partitions of interest are
\begin{equation}
 \label{partitionsm}
{\mathcal I}_m \cup {\mathcal J}_m = \{ i_1, \ldots, i_{g+1-2m}\}\cup \{ j_1, \ldots, j_{g+1+2m}\},
\end{equation}
where $m$ is any integer between $0$ and $\left[\frac{g+1}{2}\right]$. The corresponding characteristic $[\varepsilon_m]$ is defined by the vector
\begin{equation}
\boldsymbol{\Delta}_m = \sum_{k=1}^{g+1-2m}\widetilde{\boldsymbol{\mathfrak{A}}}_{i_k}
 +\vek{ K}_{\infty} =: \vek{\varepsilon}_m +\tau \vek{\varepsilon}'_m \, ,
\label{sumchar}
\end{equation}
where $\boldsymbol{K}_{\infty}\in \widetilde{\mathrm{Jac}}(X_g)$ is the vector of Riemann constants with base point $\infty$, which will always be used in the argument of the $\theta$--functions, and which is given as a vector in $\widetilde{\mathrm{Jac}}(X_g)$ by
\begin{equation}
\vek{ K}_\infty := \sum_{\text{all odd}\; [\mathfrak{A}_j]} \widetilde{\boldsymbol{\mathfrak A}}_j
\label{rvectorgen}
\end{equation}
(see e.g. \cite{fk80}, p. 305, for a proof).

It can be seen that characteristics with even $m$ are even, and with odd $m$ are odd. There are $\frac{1}{2}{2g+2 \choose g+1}$ different partitions with $m=0$, ${2g+2 \choose g-1}$ different partitions with $m=1$, and, in general, ${2g+2 \choose g+1-2m}$ down to ${2g+2 \choose 1}=2g+2$ partitions if $g$ is even and $m=g/2$, or ${2g+2 \choose 0}=1$ partitions if $g$ is odd and $m=(g+1)/2$.  One may check that the total number of even (odd) characteristics is indeed $2^{2g-1}\pm 2^{g-1}$. According to the Riemann theorem on the zeros of $\theta$--functions~\cite{fa73}, $\theta(\boldsymbol{\Delta}_m+\vek{v})$ vanishes to order $m$ at $\vek{v}=0$ and in particular, the function $\theta(\boldsymbol{K}_{\infty}+\vek{v})$ vanishes to order $\left[\frac{g+1}{2}\right]$ at $\vek{v}=0$.

Let us demonstrate, following \cite{fk80}, p.~303, how the set of characteristics $[\boldsymbol{\mathfrak{A}}_k] \equiv [\widetilde{\boldsymbol{\mathfrak{A}}}_k]$, $k=1,\ldots, 2g+2$ looks like in the homology basis shown in Figure 1.
 Using the notation $\vek{f}_k = \frac{1}{2}(\delta_{1k}, \ldots, \delta_{gk})^t$ and $\boldsymbol{\tau}_k$
for the $k$-th column vector of the matrix $\tau$, we find
\begin{align}
\widetilde{\boldsymbol{\mathfrak A}}_{2g+1} & = \widetilde{\boldsymbol{\mathfrak A}}_{2g+2} - \sum_{k=1}^g\int\limits_{e_{2k-1}}^{e_{2k}}\d \vek{v} = \sum_{k=1}^g\vek{f}_k, &  \to  & & [{\boldsymbol{\mathfrak A}}_{2g+1}] & = \frac12 \begin{pmatrix} 0 & 0 & \ldots & 0 & 0 \\ 1 & 1 & \ldots & 1 & 1 \end{pmatrix}, \nonumber \\
\widetilde{\boldsymbol{\mathfrak A}}_{2g} & = \widetilde{\boldsymbol{\mathfrak A}}_{2g+1} - \int\limits_{e_{2g+1}}^{e_{2g}}\d \vek{v} = \sum_{k=1}^g\vek{f}_k + \vek{\tau}_g, &  \to  & & [{\boldsymbol{\mathfrak A}}_{2g}] & = \frac12 \begin{pmatrix} 0 & 0 & \ldots & 0 & 1 \\ 1 & 1 & \ldots & 1 & 1 \end{pmatrix}, \\
\widetilde{\boldsymbol{\mathfrak A}}_{2g-1} & = \widetilde{\boldsymbol{\mathfrak A}}_{2g} - \int\limits_{e_{2k-1}}^{e_{2k}}\d \vek{v} = \sum_{k=1}^g\vek{f}_k + \vek{\tau}_g, &  \to  & & [{\boldsymbol{\mathfrak A}}_{2g-1}] & = \frac12 \begin{pmatrix} 0 & 0 & \ldots & 0 & 1 \\ 1 & 1 & \ldots & 1 & 0 \end{pmatrix} \, . \nonumber
\end{align}
Continuing in the same manner, we get for arbitrary $1 \leq k < g$
\begin{align}
\begin{split}
[{\boldsymbol{\mathfrak A}}_{2k+2}] & = \frac12 \Biggl(\overbrace{\begin{matrix} 0 & 0 & \ldots & 0 \\ 1 & 1 & \ldots & 1 \end{matrix}}^{k} \;\;
\begin{matrix} 1 & 0 & \ldots & 0 \\ 1 & 0 & \ldots & 0 \end{matrix}\Biggr), \\
[{\boldsymbol{\mathfrak A}}_{2k+1}] & = \frac12 \Biggl(\overbrace{\begin{matrix} 0 & 0 & \ldots & 0 \\ 1 & 1 & \ldots & 1 \end{matrix}}^{k} \;\;
\begin{matrix} 1 & 0 & \ldots & 0 \\ 0 & 0 & \ldots & 0 \end{matrix}\Biggr)
\end{split}
\end{align}
and finally
\begin{equation}
[\boldsymbol{{\mathfrak A}}_2] = \frac12 \begin{pmatrix} 1 & 0 & \ldots & 0 \\ 1 & 0 & \ldots & 0 \end{pmatrix} \, , \qquad
[\boldsymbol{{\mathfrak A}}_1] = \frac12 \begin{pmatrix} 1 & 0 & \ldots & 0 \\ 0 & 0 & \ldots & 0
\end{pmatrix}\, .
\end{equation}
\weglassen{ For example, we have for the homology basis
 drawn on the Figure \ref{figure-1}:
  \begin{align*}
&[{\mathfrak
A}_{1\phantom{0}}]=\frac12\left[{}_1^0{}_0^0{}_0^0{}_0^0{}_0^0\right],\quad
[{\mathfrak
A}_{2\phantom{0}}]=\frac12\left[{}_1^1{}_0^0{}_0^0{}_0^0{}_0^0\right],\quad
[{\mathfrak
A}_{3\phantom{0}}]=\frac12\left[{}_0^1{}_1^0{}_0^0{}_0^0{}_0^0\right],\\
&[{\mathfrak
A}_{4\phantom{0}}]=\frac12\left[{}_0^1{}_1^1{}_0^0{}_0^0{}_0^0\right],\quad
[{\mathfrak
A}_{5\phantom{0}}]=\frac12\left[{}_0^1{}_0^1{}_1^0{}_0^0{}_0^0\right],\quad
[{\mathfrak
A}_{6\phantom{0}}]=\frac12\left[{}_0^1{}_0^1{}_1^1{}_0^0{}_0^0\right],\\
&[{\mathfrak
A}_{7\phantom{0}}]=\frac12\left[{}_0^1{}_0^1{}_0^1{}_1^0{}_0^0\right],\quad
[{\mathfrak A}_{8\phantom{0}}]
=\frac12\left[{}_0^1{}_0^1{}_0^1{}_1^1{}_0^0\right],\quad
[{\mathfrak
A}_{9\phantom{0}}]=\frac12\left[{}_0^1{}_0^1{}_0^1{}_0^1{}_1^0\right],\\
&[{\mathfrak
A}_{10}]=\frac12\left[{}_0^1{}_0^1{}_0^1{}_0^1{}_1^1\right],\quad
[{\mathfrak
A}_{11}]=\frac12\left[{}_0^1{}_0^1{}_0^1{}_0^1{}_0^1\right],\quad
[{\mathfrak
A}_{12}]=\frac12\left[{}_0^0{}_0^0{}_0^0{}_0^0{}_0^0\right].
\end{align*}
} 
The characteristics with even indices, corresponding to the branch points $e_{2n}$, $n=1,\ldots, g$, are odd (except for $[{\mathfrak A}_{2g+2}]$ which is zero); the others are even.
Therefore in the basis drawn in Figure \ref{figure-1} we get
\begin{equation}
\vek{K}_\infty =  \sum_{k=1  }^g \widetilde{\boldsymbol{\mathfrak A}}_{2k} \, .
\label{rvector}
\end{equation}
The formula (\ref{rvector}) is in accordance with the classical theory where the vector of Riemann constants is defined as (see Fay \cite{fa73}, Eq. (14))
\begin{equation}
\mathrm{Divisor} \, \boldsymbol{K}_{P_0} = \Delta-(g-1)P_0 \, ,
\end{equation}
where $\Delta$ is divisor of degree $g-1$ that is the {\em Riemann divisor}. In the case considered $P_0=\infty$ and $\Delta = e_{2} + e_4 + \ldots + e_{2g} -\infty$. The calculation of the divisor of the differential $\prod_{k=1}^g(x-e_{2k}) \mathrm{d}x/y$ leads to required conclusion $2 \Delta = \mathcal{K}_{X_g}$ where  $\mathcal{K}_{X_g}$ is canonical class.

\subsection{$\sigma$--functions}

The Kleinian $\sigma$--function of the hyperelliptic curve $X_g$ is defined over the Jacobian $\mathrm{Jac}(X_g)$ as
\begin{equation}
\sigma(\boldsymbol{u};M) := C\theta[\boldsymbol{K}_{\infty}]( (2\omega)^{-1} \boldsymbol{u};\tau   ) \mathrm{exp}\left\{  \boldsymbol{u}^T \varkappa \boldsymbol{u}  \right\} \label{sigma} \, ,
\end{equation}
where $\varkappa=\eta (2\omega)^{-1}$, the constant
\begin{equation}
C = \sqrt{ \frac{\pi^g}{\mathrm{det}(2\omega)} }\left(\prod_{1\leq i <j \leq 2g+1} (e_i-e_j)\right)^{-1/4} \,, \label{sigmac}
\end{equation}
and $M$ defined in \eqref{defM} contains the set of all moduli $2\omega,2\omega'$ and $2\eta, 2\eta'$. In what follows we will use the shorter notation $\sigma(\boldsymbol{u};M)=\sigma(\boldsymbol{u})$. Sometimes the $\sigma$--function (\ref{sigma}) is called fundamental $\sigma$--function.

The multivariable $\sigma$--function (\ref{sigma}) represents a natural generalization of the Weierstra{\ss}
$\sigma$--function given by
\begin{equation}
\sigma(u)= \sqrt{\frac{\pi}{2\omega}} \frac{\epsilon}{\sqrt[4]{(e_1-e_2)(e_1-e_3)(e_2-e_3)}} \vartheta_1\left( \frac{u}{2\omega}\right) \mathrm{exp} \left\{ \frac{\eta u^2}{2\omega} \right\},\quad \epsilon^8=1\, , \label{fundamental}
\end{equation}
where $\vartheta_1$ is the standard $\theta$--function.

The fundamental $\sigma$--function introduced by the formula (\ref{sigma}) respects the following properties
\begin{itemize}
\item it is an entire function on  $\mathrm{Jac}(X_g)$,
\item  it satisfies the two sets of functional equations
\begin{equation}
\begin{split}
\sigma(\boldsymbol{u}+2\omega \boldsymbol{k}+2\omega'\boldsymbol{k}; M) & = e^{2 (\eta\boldsymbol{k}+\eta'\boldsymbol{k}') (\boldsymbol{u}+\omega\boldsymbol{k}+\omega'\boldsymbol{k}')} \sigma(\boldsymbol{u};M)\\
\sigma(\boldsymbol{u};(\gamma M^T)^T) & = \sigma(\boldsymbol{u};M) \, , \end{split}
\end{equation}
where $\gamma \in \mathrm{Sp}(2g,\mathbb Z)$, that is, $\gamma J \gamma^{T} = J$, and $M^T$ is the matrix $M$ with interchanged submatrices $\omega^\prime$ and $\eta$.
The first of these equations displays the {\it periodicity property}, and the second one the {\it modular property}. 

\item In the vicinity of the origin the power series of $\sigma(\boldsymbol{u})$ is of the form
\begin{equation}
\sigma(\boldsymbol{u}) = S_{\boldsymbol{\pi}}(\boldsymbol{u})+\text{higher order terms} \, ,  \label{sigmaseries}
\end{equation}
where $S_{\boldsymbol{\pi}}(\boldsymbol{u})$ are the Schur--Weierstra{\ss} functions whose definition we will recall in the next subsection.
\end{itemize}

\subsection{Schur--Weierstra{\ss} functions}

The Schur function is a polynomial in the variables $u_1,u_2,\ldots$ built by a partition
$\boldsymbol{\lambda}: \lambda_1\geq \lambda_2\geq \ldots \geq
\lambda_n$ of weight $|\boldsymbol{\lambda}|=\sum_{i=1}^n\lambda_i$. Any partition $\boldsymbol{\lambda}$ can be written in the Frobenius notation $\boldsymbol{\lambda}=(\alpha_1,\ldots,\alpha_r\vert \beta_1,\ldots,\beta_r)$, where the number $r$ is the rank of the partition and the integers $ (\alpha_j,\beta_j) $ are the numbers of nodes in the Young diagram to the left from $j$-th diagonal node and down to it (see \cite{sag01}).


In what follows we will deal with the special kind of Schur functions that are related to the hyperelliptic curve and that we will call Schur--Weierstra{\ss} functions $S_{\boldsymbol{\pi}}(\boldsymbol{u})$  following \cite{BEL99}.
The associated partition $\boldsymbol{\pi}$ is defined by the Weierstra{\ss} gap sequence $\mathrm{w}=(w_1,\ldots,w_g)$ at the infinite branch point, $\mathrm{w}=(1,3,\ldots,2g-1)$ by the formula
\begin{equation}
\pi_i=w_{g-k+1}+k-g, \quad i=1,\ldots,g\label{part} \,.
\end{equation}
In the considered case the associated Young diagrams are therefore symmetric and satisfy the constraint $\pi_{k}-\pi_{k+1}=1$, i.e., in the cases $g=5$ and $g=6$ we have the diagrams
 \[ \yng(5,4,3,2,1)\quad\text{and}\quad  \yng(6,5,4,3,2,1) \]
corresponding to a partition of rank $3$. In general, for arbitrary $g$ the rank of the partition is the integer part $\left[\frac{g+1}{2}\right]$.

For the polynomials $S_{\boldsymbol{\pi}}(\boldsymbol{u})$ the following representation is valid
\begin{equation}
S_{\boldsymbol{\pi}}=\mathrm{det} \left( c_{\pi_i-i+j} \right)_{1\leq i,j \leq g}
\end{equation}
where $e_k$ is given by the determinant
\begin{equation}
c_k=\frac{1}{k!}\left|\begin{array}{ccccc} p_1&1&0&\ldots&0\\
     p_2&p_1&2&\ldots&0\\
     \ldots&\ldots&\ldots&\ldots&\\
     p_{k-1}&p_{k-2}&p_{k-3}&\ldots&k-1\\
     p_k&p_{k-1}&p_{k-2}&\ldots&p_1                  \end{array} \right|
\end{equation}
and the quantities $p_k$ are related to the Jacobian variables $(u_1,\ldots,u_g)$ as
\begin{equation}
p_k=k u_{ g-[k/2 ]}, \quad k=1, \ldots, 2g \, .
\end{equation}

For example,
\begin{align}
g & = 1: & \quad S_{1}(u_1) & = u_1,\label{schur1} \\
g & = 2: & \quad S_{2,1}(u_1,u_2) & = \frac13 u_2^3-u_1 , \label{schur2} \\
g & = 3: & \quad  S_{3,2,1}(u_1,u_2,u_3) & = \frac{1}{45}u_3^6 - \frac13 u_2 u_3^3 - u_2^2+u_1u_3,\label{schur3}\\
g & = 4: & \quad S_{4,3,2,1}(u_1,u_2,u_3,u_4) & = \frac{1}{4725} u_4^{10} - \frac{1}{105} u_4^7 u_3 + \frac{1}{15} u_2u_4^5 - u_4 u_3^3 - \frac{1}{3} u_4^3 u_1 \label{schur4}\\
& & & \qquad + u_2 u_3 u_4^2 - u_2^2 + u_1 u_3 \, . \nonumber
\end{align}

\subsection{Kleinian $\wp$--functions}

The $\wp$--functions are a natural generalization of the corresponding Weierstra{\ss} functions and given as logarithmic derivatives of $\sigma$
\begin{align}
\begin{split}
\wp_{ij}(\boldsymbol{u}) & = -\frac{\partial^2}{ \partial u_i\partial u_j} \;\mathrm{ln}\,\sigma(\boldsymbol{u}), \\
\wp_{ijk}(\boldsymbol{u}) & = -\frac{\partial^3}{ \partial u_i\partial u_j\partial u_k} \;\mathrm{ln}\,\sigma(\boldsymbol{u}) \,, \quad \text{etc.,}
\end{split}
\end{align}
where $i,j,k \in \{1,\ldots,g\}$. In this notation the Weierstra{\ss} $\wp$--function is $\wp_{1,1}(u)$. For convenience, we will also denote the derivatives of the $\sigma$--function by
\begin{equation}
\sigma_i(\boldsymbol{u})=\frac{\partial}{\partial u_i} \sigma(\boldsymbol{u}),\quad \sigma_{ij}(\boldsymbol{u})=\frac{\partial^2}{\partial u_i\partial u_j} \sigma(\boldsymbol{u}), \qquad \text{etc.}
\end{equation}

The Jacobi inversion problem is the problem of the inversion of the Abel map and can be formulated as follows: for an arbitrary vector $\boldsymbol{u} \in \mathrm{Jac}(X_g)$ find the symmetric functions of $g$ points $P_1, \ldots, P_g \in X_g$ from the equation (\ref{jip}), that is  $\boldsymbol{u}=\sum_{k=1}^g \int_\infty^{P_k} d\boldsymbol{u}$. In the considered case of a hyperelliptic curve Jacobi's inversion problem is written in coordinate notation as
\begin{align}\begin{split}
\int_{P_0}^{P_1} \frac{\mathrm{d}x}{y}+\ldots+\int_{P_0}^{P_g} \frac{\mathrm{d}x}{y} & = u_1 \,,\\
\int_{P_0}^{P_1} \frac{x\mathrm{d}x}{y}+\ldots+\int_{P_0}^{P_g} \frac{x\mathrm{d}x}{y} & = u_2 \,,\\
 & \quad \vdots \\
\int_{P_0}^{P_1} \frac{x^{g-1}\mathrm{d}x}{y}+\ldots+\int_{P_0}^{P_g} \frac{x^{g-1}\mathrm{d}x}{y} & =u_g \, ,
\end{split}\label{JIP2}
\end{align}
where $P_k = (x_k,y_k)$. It can be solved in terms of Kleinian $\wp$--functions, i.e., $x_1,\ldots,x_g$ are given by the $g$ solutions of
\begin{align}\begin{split}
& x^g - \wp_{gg}(\boldsymbol{u}) x^{g-1} -\ldots - \wp_{g,1}(\boldsymbol{u})=0 \,,\\
\text{and} \quad & y_k = -\wp_{ggg}(\boldsymbol{u})x_k^{g-1}-\ldots - \wp_{gg1}(\boldsymbol{u}), \quad k=1,\ldots,g \, .
\end{split} \label{JIPGEN}
\end{align}
Another expressions for symmetric functions $\sum_{k=1}^g x_k^l$, with $l = 1, 2, 3, \ldots$, are given in \cite{vanha95}.

Among the various differential relations between the Kleinian $\wp$--functions we quote the representation of the Jacobi variety as algebraic variety in $\mathbb{C}^{ g+\frac{g(g+1)}{2} }$ obtained in \cite{BEL97} and used in the foregoing development. It is described by the set of cubic relations that can also be represented as minors of a  certain matrix \cite{BEL97},
\begin{eqnarray}
\wp_{ggi}\wp_{ggk} & = & 4\wp_{gg}\wp_{gi}\wp_{gk}-2(\wp_{gi}\wp_{g-1,k}+ \wp_{gk}\wp_{g-1,i})+4(\wp_{gk}\wp_{g,i-1}+\wp_{gi}\wp_{g,k-1}) \nonumber \\
& & + 4\wp_{k-1,i-1}-2(\wp_{k,i-2}+\wp_{i,k-2})+\lambda_{2g}\wp_{gk}\wp_{gi}
+\frac{\lambda_{2g-1}}{2}(\delta_{ig}\wp_{gk}+\delta_{kg}\wp_{gi} ) \label{cubic} \\
& & + \lambda_{2i-2} \delta_{ik} + \frac12(\lambda_{2i-1} \delta_{k,i+1} + \lambda_{2k-1} \delta_{i,k+1}), \quad 1\leq i,k \leq g \, . \nonumber
\end{eqnarray}
For $g=1$ Eq.~(\ref{cubic}) reduces to the Weierstra{\ss} cubic ${\wp'}^2=4\wp^3-g_2\wp -g_3$ if we set $\lambda_2=0$, $\wp_{111}=\wp', \wp_{11}=\wp$.

\subsection{Period matrices}

The construction of the fundamental $\sigma$--functions requires the knowledge of the first period matrix $(2\omega, 2\omega')$ {\it and} the second period matrix $(2\eta,2\eta')$ which satisfy the generalized Legendre relation (\ref{Legendre}). We will show that the second period matrix can be constructed in terms of the first period matrix and even $\theta$-constants.

For that purpose we consider any even nonsingular half-period of the $\theta$--function $(2\omega)^{-1} \boldsymbol{\mathfrak A}_{\mathcal{I}_0} + \boldsymbol{K_{\infty}}$ where the half-period
\begin{equation}
\boldsymbol{\mathfrak A}_{\mathcal{I}_0}=\int_{\infty}^{e_{i_1}}\mathrm{d}\boldsymbol{u}+\ldots
+\int_{\infty}^{e_{i_g}}\mathrm{d}\boldsymbol{u}
\end{equation}
corresponds to the partition of the branch points
\begin{equation}
\{i_1,\ldots, i_g,2g+2\}  \cup   \{j_1,\ldots, j_{g+1}\}=\mathcal{I}_0\cup\mathcal{J}_0 \weglassen{\{1,\ldots, 2g+1\}} \,.
\end{equation}
Then from (\ref{JIPGEN}) with $x_k = e_{i_k}$, $\boldsymbol{u}=\boldsymbol{\mathfrak A}_{\mathcal{I}_0}$, follows an expressions for $\wp_{ig}(\boldsymbol{\mathfrak A}_{\mathcal{I}_0})$ in terms of symmetric functions of the elements $e_{i_k}$, $i_k\in \mathcal{I}_0$,
\begin{align}\begin{split}
e_{i_1}+\ldots+e_{i_g} & = \wp_{gg}(\boldsymbol{\mathfrak A}_{\mathcal{I}_0}), \\
e_{i_1} e_{i_2} + \ldots + e_{i_{g-1}}e_{i_g} & =-\wp_{g-1,g}(\boldsymbol{\mathfrak A}_{\mathcal{I}_0}), \\
& \vdots\\
e_{i_1} \cdots e_{i_g} &= (-1)^{g-1}\wp_{1g}(\boldsymbol{\mathfrak A}_{\mathcal{I}_0}) \,.
\end{split}
\label{JIPGENE}
\end{align}
One can see that all two-index symbols $\wp_{ij}(\boldsymbol{\mathfrak A}_{\mathcal{I}_0})$ can be expressed in terms of symmetric functions of two sets of variables $e_{i}, \; i\in \mathcal{I}_0$ and $e_{j}, \; j\in \mathcal{J}_0$. That follows from the fundamental cubic relation (\ref{cubic}). Indeed $\wp_{igg}(\boldsymbol{\mathfrak A}_{\mathcal{I}_0})=0$ for all $i=1,\ldots,g$ and we get $\frac{g(g+1)}{2}$ relations to which (\ref{JIPGENE}) are substituted as well as expressions for the $\lambda's$ in terms of $e_i$, $i=1,\ldots, 2g+1$. These equations can always be solved with respect of the remaining $\wp_{jk}(\boldsymbol{\mathfrak A}_{\mathcal{I}_0})$. Therefore, all $\wp_{i,j}(\boldsymbol{\mathfrak A}_{\mathcal{I}_0})$ are known in terms of the branch points $e_k$.

\begin{proposition} \label{kappaprop}
Let $\boldsymbol{\mathfrak A}_{\mathcal{I}_0}+2\omega \boldsymbol{K}_{\infty}$ be an arbitrary even nonsingular half-period corresponding to the $g$ branch points of the set of indices $\mathcal{I}_0=\{i_1,\ldots,i_g\}$.
We define the symmetric $g\times g$ matrices
\begin{equation}
\mathfrak{P}(\boldsymbol{\mathfrak A}_{\mathcal{I}_0}) := \left(\wp_{ij}(\boldsymbol{\mathfrak A}_{\mathcal{I}_0})   \right)_{i,j=1,\ldots,g}
\label{matrixP}
\end{equation}
and
\begin{equation}
\mathfrak{T}(\boldsymbol{\mathfrak A}_{\mathcal{I}_0}) := \left( - \frac{\partial^2}{ \partial z_i\partial z_j}\,\mathrm{log}\,\left. \theta[\boldsymbol{K}_{\infty}](\boldsymbol{z};\tau)
\right|_{\boldsymbol{z}=(2\omega)^{-1}\boldsymbol{\mathfrak A}_{\mathcal{I}_0}} \right)_{i,j=1,\ldots,g} \,.
\end{equation}
Then the $\varkappa$-matrix is given by
\begin{equation}
\varkappa = - \frac12\mathfrak{P}(\boldsymbol{\mathfrak A}_{\mathcal{I}_0})-\frac12
((2\omega)^{-1})^T \mathfrak{T}(\boldsymbol{\mathfrak A}_{\mathcal{I}_0}) (2\omega)^{-1} \label{kappa}
\end{equation}
and the half-periods $\eta$ and $\eta'$ of the meromorphic differentials can be represented as
\begin{equation}
\eta = 2 \varkappa \omega, \qquad \eta' = 2 \varkappa \omega' - \frac{\imath\pi}{2}(\omega^{-1})^T \,.
\end{equation}
\end{proposition}

We remark that (\ref{kappa}) represents the natural generalization of the Weierstra{\ss} formulae
\begin{equation}
2\eta\omega=-2e_1\omega^2-\frac12 \frac{\vartheta_2''(0)}{\vartheta_2(0)}\,,\quad
2\eta\omega=-2e_2\omega^2-\frac12 \frac{\vartheta_3''(0)}{\vartheta_3(0)}\,,\quad
2\eta\omega=-2e_3\omega^2-\frac12 \frac{\vartheta_4''(0)}{\vartheta_4(0)}\, ,
\end{equation}
see e.g. the Weierstra{\ss}--Schwarz lectures, \cite{weierstrass893} p. 44.
Therefore Proposition \ref{kappaprop} allows the reduction of the variety of moduli necessary for the calculation of the $\sigma$-- and $\wp$--functions to the first period matrix. The generalization of the result (\ref{kappa}) to non--hyperelliptic curves is considered in \cite{enko11}.

\section{Inversion of one hyperelliptic integral}\label{sec:inversionhei}

Classically it is known that only symmetric functions of $g$ points of algebraic curves of genus $g$ can be presented as single--valued functions over the Jacobi variety of the curve. Nevertheless, one hyperelliptic integral can also be inverted analytically in terms of $\sigma$--functions restricted to the $\theta$--divisor.

\subsection{Stratification of the $\theta$--divisor}

The $\theta$--divisor $\widetilde{\Theta}$ is defined as the subset of $\widetilde{\mathrm{Jac}}(X_g)$ that nullifies $\theta$ and, therefore, the $\sigma$--function, i.e.
\begin{equation}
\widetilde{\Theta}=\left\{ \boldsymbol{v}\in \widetilde{\mathrm{Jac}}(X_g) \, \vert \,\theta(\boldsymbol{v})\equiv 0 \right\} \,.
\label{Theta1}
\end{equation}
The subset $\widetilde{\Theta}_k\subset \widetilde{\Theta}$, $0 \leq k < g$, is called $k$-th stratum if each point $\boldsymbol{v}\in \widetilde{\Theta}$ admits a parametrization
\begin{equation}
\widetilde{\Theta}_k := \Biggl\{\boldsymbol{v} \in \widetilde{\Theta} \Big|
\boldsymbol{v} = \sum_{j=1}^{k}\int_{\infty}^{P_j} \mathrm{d}\boldsymbol{v} + \boldsymbol{K}_{\infty}\Biggr\} \, , \label{strata}
\end{equation}
where $\widetilde{\Theta}_0=\{\boldsymbol{K}_{\infty}\}$ and $\widetilde{\Theta}_{g-1} = \widetilde{\Theta}$. We furthermore denote $\widetilde{\Theta}_g= \widetilde{\mathrm{Jac}}(X_g)$. The following natural embedding is valid:
\begin{equation}
\label{stratification} \widetilde{\Theta}_0\;\subset\; \widetilde{\Theta}_1\; \subset \ldots \subset\; \widetilde{\Theta}_{g-1}\; \subset \; \widetilde{\Theta}_g = \widetilde{\mathrm{Jac}}(X_g) \,.
\end{equation}

We define the $\theta$--function to be {\it vanishing to the order} $m(\widetilde{\Theta}_k)$ along the stratum $\widetilde{\Theta}_k$ if for all sets $\alpha_j$, $j = 1, \ldots, g$ with $0 \leq \alpha_1 + \ldots + \alpha_g < m$
\begin{equation}
\frac{\partial^{\alpha_1+\ldots+\alpha_g}}{\partial u_1^{\alpha_1} \ldots \partial u_g^{\alpha_g} }\theta(\boldsymbol{v\vert\tau})\equiv 0,\quad \forall \boldsymbol{v}\in \widetilde{\Theta}_k \, , \label{vanish}
\end{equation}
while there is a certain set of $\alpha_j$, with $\alpha_1 + \ldots + \alpha_g = m$ so that (\ref{vanish}) does not hold. The orders $m(\widetilde{\Theta}_k)$ of the vanishing of $\theta(\widetilde{\Theta}_k + \boldsymbol{v})$ along the stratum $\widetilde{\Theta}_k$ for the first genera are given in Table~\ref{table1}.

\begin{table}[t]
\begin{tabular}{ |l|l|l|l|l|l|l|l|l| }
$g$&$m(\widetilde{\Theta}_0)$&$m(\widetilde{\Theta}_1)$&$m(\widetilde{\Theta}_2)$&
$m(\widetilde{\Theta}_3)$&$m(\widetilde{\Theta}_4)$&$m(\widetilde{\Theta}_5)$&$m(\widetilde{\Theta}_6)$\\
\hline
1&1&0&-&-&-&-&-\\
2&1&1&0&-&-&-&-\\
3&2&1&1&0&-&-&-\\
4&2&2&1&1&0&-&-\\
5&3&2&2&1&1&0&-\\
6&3&3&2&2&1&1&0
\end{tabular}
\vskip0.3cm
\caption{Orders $m(\widetilde{\Theta}_k)$ of zeros $\theta(\widetilde{\Theta}_k+\boldsymbol{v})$ at $\boldsymbol{v}=0$ on the strata $\widetilde{\Theta}_k$.} \label{table1}
\end{table}

In the following we focus on the stratum $\widetilde{\Theta}_1$ corresponding to the variety $\Theta_1 \subset \mathrm{Jac}(X_g)$, which is the image of the curve inside the Jacobian,
\begin{equation}
\Theta_1 = \left\{\boldsymbol{u} \in \Theta \Bigg|  \boldsymbol{u}=\int_{\infty}^{P}\mathrm{d}\boldsymbol{u}\right\} \,,\quad \text{and}\quad
\widetilde{\Theta}_1=(2\omega)^{-1}\Theta_1+\boldsymbol{K}_{\infty} \, .
\end{equation}

We remark that another stratification was introduced in \cite{vanha95} for hyperelliptic curves of even order with two infinite points $\infty_+$ and $\infty_-$ that was implemented for studying the poles of function on Jacobians of these curves. The same problem relevant to strata of the $\theta$--divisor was studied in \cite{abendfed00}.

\subsection{Inversion formulae}

For the case of genus two the inversion of a holomorphic hyperelliptic integral by the method of restriction to the $\theta$--divisor was obtained independently by Grant \cite{gr90} and Jorgenson \cite{jo92} in the form
\begin{equation}
x=-\left.\frac{\sigma_1(\boldsymbol{u})}{\sigma_2(\boldsymbol{u})}\right|_{\sigma(\boldsymbol{u})=0}, \qquad \boldsymbol{u}=(u_1,u_2)^T\label{gr} \,.
\end{equation}
This result was implemented in \cite{EPR03}, and explicitely worked out in the series of publications \cite{HackmannLaemmerzahl08,HackmannLaemmerzahl08a,Hackmannetal08,Hackmannetal09,Hackmannetal010}, and others.

The case of genus three was studied by \^Onishi \cite{on98}, where the inversion formula is given in the form
\begin{equation}
x=-\left.\frac{\sigma_{13}(\boldsymbol{u})}{\sigma_{23}(\boldsymbol{u})}\right|_{\sigma(\boldsymbol{u})=\sigma_3(\boldsymbol{u})=0}, \qquad \boldsymbol{u}=(u_1,u_2,u_3)^T\label{onishi} \,.
\end{equation}
Formula (\ref{onishi}) is based on the detailed analysis of the genus three KdV hierarchy and its restriction to the $\theta$--divisor. Below we will present the generalization of (\ref{gr}) and (\ref{onishi}) to higher genera. For doing so we first analyze the Schur--Weierstra{\ss} polynomials that represent the first term of the expansion of $\sigma(\boldsymbol{u})$ in the vicinity of the origin $\boldsymbol{u}\sim 0$.

\subsection{Schur--Weierstra{\ss} polynomials on $\Theta_k$-strata}

The Schur--Weierstra{\ss} polynomials associated to the curve $X_g$ are defined in the space $\mathbb{C}^g\ni(u_1,\ldots,u_g)$ by the Weierstra{\ss} gap sequence at the infinite branch point that for $g>1$ is always a Weierstra{\ss} point. The $\theta$--divisor $\Theta$ and its strata $\Theta_k$ in the vicinity of the origin $\boldsymbol{u}\sim 0$ are given as polynomials in $\boldsymbol{u}$. An analysis of the Schur--Weierstra{\ss} polynomials leads to the following
\begin{proposition}\label{SWPOL}
The following statements are valid for the Schur--Weierstra{\ss} polynomials $S_{\boldsymbol{\pi}}(\boldsymbol{u})$
associated with a partition $\boldsymbol{\pi} \in \Theta_1$:
\begin{enumerate}

\item 
In the vicinity of the origin, an element $\boldsymbol{u}$ of the first stratum $\Theta_1 \subset \Theta$ is singled out by
\begin{equation}
S_{\boldsymbol{\pi}}(\boldsymbol{u})=0,\quad \frac{\partial^j }{\partial u_g^j}  S_{\boldsymbol{\pi}}(\boldsymbol{u})=0 \quad \forall\,j=1,\ldots,g-2 \,.
\end{equation}

\item The derivatives fulfill
\begin{equation}
\frac{\partial^j }{\partial u_g^j}  S_{\boldsymbol{\pi}}(\boldsymbol{u})
\begin{cases}
\equiv 0 \quad \text{if} \quad   1\leq j < \frac{g(g-1)}{2}\\
\not\equiv 0 \quad \text{if} \quad   j \geq \frac{g(g-1)}{2}
\end{cases} \text{with}\quad \boldsymbol{u}\in {\Theta}_1 \, .
\end{equation}

\item The following equalities are valid for $\boldsymbol{u}\in {\Theta}_1$
\begin{equation}
x\cong-\frac{1}{u_g^2}=- \frac{
 \dfrac{\partial^{M}}{\partial u_1 \partial u_{g}^{M-1} }  S_{\boldsymbol{\pi}}(\boldsymbol{u})}
{\dfrac{\partial^{M}}{\partial u_2\partial u_{g}^{M-1} } S_{\boldsymbol{\pi}}(\boldsymbol{u})}=-
\frac{
 \dfrac{\partial^{M+1}}{\partial u_1 \partial u_{g}^M }  S_{\boldsymbol{\pi}}(\boldsymbol{u})}
{\dfrac{\partial^{M+1}}{\partial u_2\partial u_{g}^M } S_{\boldsymbol{\pi}}(\boldsymbol{u})}  \, ,
\end{equation}
where $M= \frac12 (g-2)(g-3) + 1$.
\item The order of vanishing of $S_{\boldsymbol{\pi}}$ restricted to $\Theta_1$ is the rank of the partition $\pi$.
\end{enumerate}
\end{proposition}

It was noted in \cite{BEL99} that the Schur--Weierstra{\ss} polynomials respect all statements of the Riemann singularity theorem. In particular, if
\begin{equation}
\boldsymbol{Z}=\left(\frac{z^{2g-1}}{2g-1},\ldots, \frac{z^{2k-1}}{2k-1}\ldots,\frac{z^3}{3},z \right)
\end{equation}
and if $\boldsymbol{\pi}$ is the partition at the infinite Weierstra{\ss} point of the hyperelliptic curve $X_g$ of genus $g$, then the function
\begin{equation}
G(z) := S_{\boldsymbol{\pi}}(\boldsymbol{Z}-\boldsymbol{u})
\end{equation}
either has $g$ zeros or vanishes identically. This result was extended in~\cite{matprev08, matprev10}.
Moreover we will conjecture here that the properties of the Schur--Weierstra{\ss} polynomials given in Proposition \ref{SWPOL} can be ``lifted" to the fundamental $\sigma$--function (\ref{sigma}).

\subsection{Inversion for higher genera}

The above analysis permits to conjecture the following inversion formula for the general case of hyperelliptic curves of genus $g>2$
\begin{equation}
x=-\left. \frac{
 \dfrac{\partial^{M+1}}{\partial u_1 \partial u_{g}^M }  \sigma(\boldsymbol{u})}
{\dfrac{\partial^{M+1}}{\partial u_2\partial u_{g}^M } \sigma(\boldsymbol{u})} \right|_{\boldsymbol{u}\in \Theta_1}, \qquad M=\frac{(g-2)(g-3)}{2}+1
\label{matprev}
\end{equation}
and
\begin{equation}
{\Theta}_1 = \left\{\boldsymbol{u} \in {\rm Jac}(X_g) \Big| \sigma(\boldsymbol{u})=0,\quad \frac{\partial^j }{\partial u_g^j} \sigma(\boldsymbol{u})=0 \quad \forall\,j=1,\ldots, g-2\right\} \,. \label{restriction}
\end{equation}

The analog of this formula for strata $\Theta_{k}$, $ 1<k<g $ and $(n,s)$-curves in the terminology of \cite{BEL99} was recently considered by Matsutani and Previato   \cite{matprev08}, \cite{matprev10}.

We remark that the half--periods associated with one of the branch points $e_1,\ldots, e_{2g+1}$ are elements of the first stratum and, therefore, the following proposition is valid:
\begin{proposition} Let the curve $X_g$ be of genus $g > 2$ and let $\boldsymbol{\mathfrak A}_i$ be the half-period that is the Abelian image with the base point $P_0=(\infty,\infty)$ of a branch point $e_i$. Then
\begin{equation}
e_i = -\frac{ \dfrac{\partial^{M+1}}{\partial u_1 \partial u_{g}^M }  \sigma(\boldsymbol{\mathfrak A}_i)}
{\dfrac{\partial^{M+1}}{\partial u_2\partial u_{g}^M } \sigma(\boldsymbol{\mathfrak A}_i)} \, ,
\label{matprev}
\end{equation}
where $M = \frac{1}{2} (g-2)(g-3) + 1$.
\end{proposition}

The formula (\ref{matprev}) can be considered as an equivalent of the Thomae formulae  \cite{thomae70}. Similar formulae can be written on other strata $\Theta_k$, but we only mention here the set of formulae that we are using in our construction.


\begin{proposition}
Let $X_g$ be a hyperelliptic curve of genus $g$ and consider a partition
\begin{equation}
 \mathcal{I}_1\cup \mathcal{J}_1= \{ i_1,\ldots, i_{g-1} \}\cup\{j_1,\ldots, j_{g+3} \}
  \label{partg1} \end{equation}
of branch points such that the half-periods
\begin{equation}
(2\omega)^{-1}\boldsymbol{\mathfrak A}_{\mathcal{I}_1} + \boldsymbol{K}_{\infty}\in \widetilde{\Theta}_{g-1}\cup \widetilde{\Theta}_{g-2}
\end{equation}
are nonsingular odd half-periods. Consider two cases:
\begin{align}
\mathcal{I}_1':= \{i_1,\ldots,i_{g-1}\} & \not\ni 2g+2 \,,\\
\mathcal{I}_1'':= \{i_1,\ldots,i_{g-1}\} & \ni 2g+2, \quad \text{i.e.}\quad i_{g-1}=2g+2 \,.
\end{align}
Denote by $s_k(\mathcal{I}_1')$ and $s_k(\mathcal{I}_1'')$ the elementary symmetric function of order $k$ built by the branch points $e_{i_1},\ldots,e_{i_{g-1}}$ and $e_{i_1},\ldots,e_{i_{g-2}}$ correspondingly. Then the following formulae are valid
\begin{align}\begin{split}
s_k(\mathcal{I}_1')&=(-1)^{k} \frac{\sigma_{g-k}} {\sigma_g} \left({\boldsymbol{\mathfrak A}_{\mathcal{I}_1'}}\right) \,,\quad k=1,\ldots,g-1 \,,\\
s_{k-1}(\mathcal{I}_1'')&=(-1)^{k-1} \frac{\sigma_{g-k}} {\sigma_{g-1}} \left({\boldsymbol{\mathfrak A}_{\mathcal{I}_1''}}\right)
 \,,\quad k=2,\ldots,g-1\,,
 \end{split}
\label{matprevform}
\end{align}
where we denoted for typographic convenience $\frac{\sigma_i(\boldsymbol{\mathfrak A})}{\sigma_j(\boldsymbol{\mathfrak A})} =: \frac{\sigma_i}{\sigma_j}(\boldsymbol{\mathfrak A})$.
\end{proposition}

The following corollary follows immediately from (\ref{matprevform}):
\begin{corollary}
Let $g>3$ and $\mathcal{I}'=\mathcal{I}''\cup\{i\}$ where  $\mathcal{I}''=\{i_1,\ldots, i_{g-2}\}$ and $i\neq 2g+2$, $i\not \in \mathcal{I}''$. Then the representation for the branch points $e_i$
\begin{equation}
e_i = -\frac{\sigma_{g-1}}{\sigma_g}\left( \boldsymbol{\mathfrak A}_{\mathcal{I}'} \right)+\frac{\sigma_{g-2}}{\sigma_{g-1}}\left( \boldsymbol{\mathfrak A}_{\mathcal{I}''} \right). \label{ei1}
\end{equation}
is valid.
\end{corollary}

The formulae (\ref{ei1}) can be understood as a generalization of those given in Bolza \cite{bolza86}. We also remark that a comparison of the two representations (\ref{matprev}) and (\ref{ei1}) of the branch point $e_i$ leads to an interesting $\theta$-constant relation.

\subsection{Calculation of moduli}\label{moduli_calc}

Calculations in terms of $\theta$-- or $\sigma$--functions are usually considered as technically cumbersome what prevents wide applications of algebro--geometric methods. In particular, the procedure of the evaluation of the period matrix in the given homology basis is technically complicated even in the case of a hyperelliptic curve. Based on the above analysis we show that modern software like the ``Maple/algcurves" package now allows the calculation of the $\theta$-- or $\sigma$--functions without drawing and even without knowledge of the homology basis, at least in the hyperelliptic case. The calculation scheme is given by the following steps:
\begin{description}
\item[Step 1.] For the given curve compute first the period matrices $(2\omega,2\omega')$ and $\tau=\omega^{-1}\omega'$ by means of the ``Maple/algcurves" code. Compute then the winding vectors, i.e., the columns of the inverse matrix \begin{equation}
(2\omega)^{-1}=(\boldsymbol{U}_1,\ldots,\boldsymbol{U}_g) \, .
\end{equation}

\item[Step 2.] We then find all nonsingular odd characteristics by direct computation of all odd $\theta$--constants. According to Table \ref{table1} we have two sets $B_1 \subset \widetilde{\Theta}_{g-1}$ and $B_2 \subset \widetilde{\Theta}_{g-2}$ of nonsingular odd half--periods. For each element of $b_1 \in B_1$ there are $e_{i_1}, \ldots, e_{i_{g-1}} \neq \infty$ such that
\begin{equation}
b_1 = \int_{\infty}^{e_{i_1}} \mathrm{d}\boldsymbol{v}+\ldots+\int_{\infty}^{e_{i_{g-1}}} \mathrm{d}\boldsymbol{v}+\boldsymbol{K}_{\infty} \in \widetilde{\Theta}_{g-1}
\end{equation}
and for each element of $b_2 \in B_2$ there are $e_{i_1}, \ldots, e_{i_{g-2}} \neq \infty$ such that
\begin{equation}
b_2 = \int_{\infty}^{e_{i_1}} \mathrm{d}\boldsymbol{v}+\ldots+\int_{\infty}^{e_{i_{g-2}}} \mathrm{d}\boldsymbol{v}+\boldsymbol{K}_{\infty} \in \widetilde{\Theta}_{g-2} \, .
\end{equation}
By using \eqref{matprevform} and the known values of the winding vectors one can find the correspondence between the sets $\{e_{i_1},\ldots, e_{i_{g-1}} \}$ and $\{e_{i_1},\ldots, e_{i_{g-2}} \}$ of branch points and the nonsingular odd characteristics $[ (2\omega)^{-1}\left(\boldsymbol{\mathfrak A}_{i_1,\ldots,i_{g-1}} \right) + \boldsymbol{K}_{\infty} ]$ and $[ (2\omega)^{-1}\left(\boldsymbol{\mathfrak A}_{i_1,\ldots,i_{g-2}} \right) + \boldsymbol{K}_{\infty}]$ . Then one can add these characteristics and find the one--to--one correspondence
\begin{equation}
\int_{\infty}^{e_{i_{g-1}}}  \mathrm{d}\boldsymbol{v}  \leftrightarrows \;[\boldsymbol{\mathfrak A}_{i_{g-1}}] ,\quad i=1,\ldots, 2g+2 \,. \label{correspondence}
\end{equation}

\item[Step 3.] Among the $2g+2$ characteristics  (\ref{correspondence}) there should be precisely $g$ odd and $g+2$ even characteristics. The sum of all odd characteristics gives the vector of Riemann constants with the base point at infinity. Check that this characteristic is of order $ \left[\frac{g+1}{2}\right]$.

\item[Step 4.] Calculate the symmetric matrix $\varkappa$ by (\ref{kappa}) and then the second period matrices $2\eta,2\eta'$ according to the Proposition \ref{kappaprop}.

\end{description}

We add two remarks:

\begin{description}
\item[Remark 1.] The proposed way to compute the $[\boldsymbol{\mathfrak{A}}_k]$ is not the only possible way. One can also use the standard Thomae formulae for the $\theta$--constants with even characteristics while we used Bolza-type formulae for odd characteristics. Also the ``Maple/algcurves" contains expressions for homology cycles written in terms of paths connecting branch points and it could be possible in principle to find the $[\boldsymbol{\mathfrak{A}}_k]$ solving the system of equations.

\item[Remark 2.]] One could imagine a case where the drawing of the homology basis is nevertheless important for the problem, e.g., in the case when the curve possesses additional symmetry. In this case one can use T.~Northower's program~\cite{nor10,nor10a} to find the symplectic transformation between the required basis and the one given by  ``Maple/algcurves".
\end{description}

\subsection{Procedure of the inversion}

We are now in the position to calculate the inversion, that is $x(t)$, of the hyperelliptic integral
\begin{equation}
\int_{\infty}^x \frac{z^k\mathrm{d}z}{\sqrt{\mathcal{P}_{2g+1}(z)}}=t \, , \qquad   0 \leq k < g \, .
\end{equation}
This will be carried through along the following steps.
\begin{description}
\item[Step 1.] We first fix the homology basis, e.g., the basis of ``Maple/algcurves" and compute all moduli of the curve according to Sec.~\ref{moduli_calc}.

\item[Step 2.] The dynamic system considered is evaluated between two branch points of the polynomial $\mathcal{P}_{2g+1}(z)$ that defines the curve (\ref{curve}). Therefore fix a branch point, say $e_i$, that is the starting point of the system evolution and find the half-period $\boldsymbol{\mathfrak A}_i = \int_\infty^{e_i} d\boldsymbol{u}$.

\item[Step 3.] Use the formula (\ref{matprev}) for $g>2$ or \eqref{gr} for $g=2$ with the argument $\boldsymbol{u}$ of the $\sigma$--function given as
\begin{equation}
\boldsymbol{u} = \boldsymbol{\mathfrak A}_i + \begin{pmatrix}  f_1(t) \\  \vdots \\ f_{k-1}(t) \\ t \\ f_{k+1}(t) \\ \vdots \\ f_g(t) \end{pmatrix} \qquad \text{with} \qquad f(0) = 0 \, ,
\end{equation}
where $f(t) = (f_1(t), \ldots, f_{g}(t))^T$ are locally given functions that resolve the conditions (\ref{restriction}) of the restriction to the stratum $\Theta_1$. The vector function $f(t)$ can be obtained from $f(0)$ using the Newton method. In this case the approximation process should be carried through for the real and imaginary parts of each function $f_i(t)$ separately.
\end{description}

We emphasize that the inversion procedure given above carries local character and the quasi-elliptic function can be defined only locally. But we believe that our method elucidates the geometric structure of the object and leads to exact calculations in contrast to numerically solving ordinary differential equations.

Below we will consider the case $g=2$ and $g=3$ and demonstrate in more detail how this scheme can be applied.

\section{Hyperelliptic curve of genus two}\label{sec:curvesgenus2}

We consider a hyperelliptic curve $X_2$ of genus two
\begin{align}
\begin{split}
w^2 & = 4(z-e_1)(z-e_2)(z-e_3)(z-e_4)(z-e_5)\\
& = 4 z^5 + \lambda_4 z^4 + \lambda_3 z^3 + \lambda_2 z^2 + \lambda_1 z + \lambda_0 \, .
\end{split} \label{curve2}
\end{align}
From (\ref{uui}) and (\ref{rrj}) the basic holomorphic and meromorphic differentials are
\begin{align}
\mathrm{d}u_1 &=\frac{\mathrm{d}z}{w}\,, & \qquad \mathrm{d}r_1 & = \frac{12 z^3+2\lambda_4z^2+\lambda_3z}{4w}\mathrm{d}z \,, \\
\mathrm{d}u_2 &=\frac{z\mathrm{d}z}{w}\,, & \qquad \mathrm{d}r_2 & = \frac{z^2}{w}\mathrm{d}z \,.
\end{align}
Then the Jacobi inversion problem for the equations
\begin{align}\begin{split}
\int_{\infty}^{(z_1,w_1)}\frac{ \mathrm{d}z}{w}+\int_{\infty}^{(z_2,w_2)}\frac{ \mathrm{d}z}{w}=u_1 \,,\\
\int_{\infty}^{(z_1,w_1)}\frac{z \mathrm{d}z}{w}+\int_{\infty}^{(z_2,w_2)}\frac{z \mathrm{d}z}{w}=u_2\end{split} \label{JIP}
\end{align}
is solved in the form
\begin{align}\begin{split}
z_1+z_2&=\wp_{22}(\boldsymbol{u}), \quad z_1z_2=-\wp_{12}(\boldsymbol{u}) \,,\\
w_k&=-\wp_{222}(\boldsymbol{u})z_k-\wp_{122}(\boldsymbol{u}), \quad k=1,2 \,.
\end{split} \label{SOLJIP}
\end{align}

\subsection{Characteristics in genus two}

The homology basis of the curve is fixed by defining the set of
half-periods corresponding to the branch points. The characteristics of the Abelian images of the branch points are defined as
\begin{equation}
[\boldsymbol{\mathfrak A}_i]=\left[\int_{\infty}^{e_i} \mathrm{d}\boldsymbol{u}\right] = \begin{pmatrix} \boldsymbol{\varepsilon}_i^{'^T} \\ \boldsymbol{\varepsilon}_i \end{pmatrix} = \begin{pmatrix} \varepsilon_{i,1}' & \varepsilon_{i,2}' \\ \varepsilon_{i,1} & \varepsilon_{i,2} \end{pmatrix} \,,
\end{equation}
what can be also written as
\[\boldsymbol{\mathfrak A}_i=2\omega  \boldsymbol{\varepsilon}_i+ 2\omega' \boldsymbol{\varepsilon'}_i, \quad i=1,\ldots,6 \,. \]

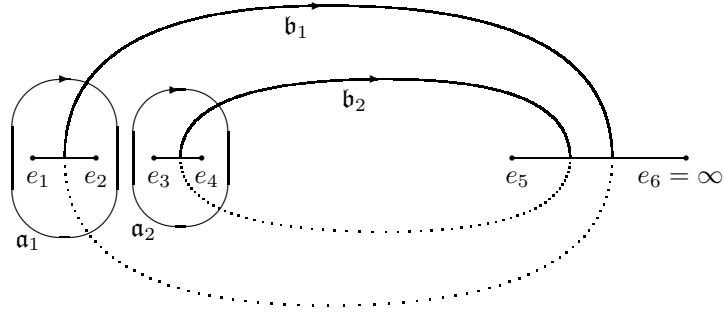
\begin{figure}
\begin{center}
\unitlength 0.7mm \linethickness{0.6pt}
\begin{picture}(150.00,80.00)
\put(9.,33.){\line(1,0){12.}} \put(9.,33.){\circle*{1}}
\put(21.,33.){\circle*{1}} \put(10.,29.){\makebox(0,0)[cc]{$e_1$}}
\put(21.,29.){\makebox(0,0)[cc]{$e_2$}}
\put(15.,33.){\oval(20,30.)}
\put(8.,17.){\makebox(0,0)[cc]{$\mathfrak{ a}_1$}}
\put(15.,48.){\vector(1,0){1.0}}
\put(32.,33.){\line(1,0){9.}} \put(32.,33.){\circle*{1}}
\put(41.,33.){\circle*{1}} \put(33.,29.){\makebox(0,0)[cc]{$e_3$}}
\put(42.,29.){\makebox(0,0)[cc]{$e_4$}}
\put(37.,33.){\oval(18.,26.)}
\put(30.,19.){\makebox(0,0)[cc]{$\mathfrak{a}_2$}}
\put(36.,46.){\vector(1,0){1.0}}
\put(100.,33.00) {\line(1,0){33.}} \put(100.,33.){\circle*{1}}
\put(133.,33.){\circle*{1}}
\put(101.,29.){\makebox(0,0)[cc]{$e_{5}$}}
\put(132.,29.){\makebox(0,0)[cc]{$e_{6}=\infty$}}
\put(59.,58.){\makebox(0,0)[cc]{$\mathfrak{b}_1$}}
\put(63.,62.){\vector(1,0){1.0}}
\bezier{484}(15.,33.00)(15.,62.)(65.,62.)
\bezier{816}(65.00,62.)(119.00,62.00)(119.00,33.00)
\bezier{35}(15.,33.00)(15.,5.)(65.,5.)
\bezier{35}(65.00,5.)(119.00,5.00)(119.00,33.00)
\put(70.,44.){\makebox(0,0)[cc]{$\mathfrak{b}_2$}}
\put(74.00,48.){\vector(1,0){1.0}}
\bezier{384}(37.,33.00)(37.,48.)(76.00,48.)
\bezier{516}(76.00,48.)(111.00,48.00)(111.00,33.00)
\bezier{30}(37.,33.00)(37.,19.)(76.00,19.)
\bezier{30}(76.00,19.)(111.00,19.00)(111.00,33.00)
\end{picture}
\end{center}
\caption{Homology basis on the Riemann surface of the curve $X_2$ with real branch points $e_1 < e_2 <\ldots < e_{6}=\infty$ (upper sheet). The cuts are drawn from $e_{2i-1}$ to $e_{2i}$, $i=1,2,3$. The $\mathfrak{b}$--cycles are completed on the lower sheet (dotted lines).} \label{figure-2}
\end{figure}

In the homology basis given in Figure \ref{figure-2} we have
\begin{align}
[\boldsymbol{\mathfrak A}_1] & = \frac{1}{2}
\begin{pmatrix} 1 & 0 \\ 0 & 0 \end{pmatrix} \, , & \quad
[\boldsymbol{\mathfrak A}_2] & = \frac{1}{2}
\begin{pmatrix} 1 & 0 \\ 1 & 0 \end{pmatrix} \, , & \quad
[\boldsymbol{\mathfrak A}_3] & = \frac{1}{2}
\begin{pmatrix} 0 & 1 \\ 1 & 0 \end{pmatrix} \\
[\boldsymbol{\mathfrak A}_4] & = \frac{1}{2}
\begin{pmatrix} 0 & 1 \\ 1 & 1 \end{pmatrix}\, , & \quad
[\boldsymbol{\mathfrak A}_5] & = \frac{1}{2}
\begin{pmatrix} 0 & 0 \\ 1 & 1 \end{pmatrix}\, , & \quad
[\boldsymbol{\mathfrak A}_6] & = \frac{1}{2}
\begin{pmatrix} 0 & 0 \\ 0 & 0 \end{pmatrix} \, .
\label{hombas2}
\end{align}
The characteristics of the vector of Riemann constants $\boldsymbol{K}_{\infty}$ is
\begin{equation}
[\boldsymbol{K}_{\infty}] = [\boldsymbol{\mathfrak{A}}_2]+ [\boldsymbol{\mathfrak{A}}_4] = \frac{1}{2} \begin{pmatrix} 1 & 1 \\ 0 & 1 \end{pmatrix}\, .
\end{equation}

From the above characteristics 16 half-periods can be build as follows: Denote the 10 half-periods for $i\neq j = 1,\ldots, 5$ that are images of two branch points as
\begin{align}
\boldsymbol{\Omega}_{i,j}=2\omega ( \boldsymbol{\varepsilon}_i+\boldsymbol{\varepsilon}_j)+
2\omega' (\boldsymbol{\varepsilon}_i'+  \boldsymbol{\varepsilon}_j') ,\quad  i = 1, \ldots, 5 \,.
\label{even}
\end{align}
Then the characteristics of the $6$ half-periods
\begin{equation}
\left[ (2\omega)^{-1} \boldsymbol{\mathfrak A}_i+\boldsymbol{K}_{\infty} \right] =: \delta_i  ,\quad i= 1, \ldots, 6
\end{equation}
are nonsingular and odd, whereas the characteristics of the $10$ half-periods
\begin{equation}
\left[(2\omega)^{-1} \boldsymbol{\Omega}_{i,j}+\boldsymbol{K}_{\infty}\right] =: \varepsilon_{i,j}, \quad 1 \leq i < j \leq 5
\end{equation}
are nonsingular and even.

Odd characteristics correspond to partitions $\{6\}\cup \{ 1,\ldots,5 \}$ and
$\{k\}\cup\{i_1,\ldots,i_4,6\}$ for $i_1,\ldots,i_4\neq k$. The first partition from these two corresponds to $\Theta_0$ and the second to $\Theta_1$.

From the solution of the Jacobi inversion problem we obtain for any $i,j =1,\ldots,5$, $i\neq j$
\begin{equation}
e_i+e_j=\wp_{22}(\boldsymbol{\Omega}_{i,j}),\quad -e_ie_j=\wp_{12}(\boldsymbol{\Omega}_{i,j})\,.
\label{JIPE1}
\end{equation}
Using the relation (see \cite{ba03}, \cite{BEL97})
\begin{equation}
\wp_{222}^2=4\wp_{22}^3+4\wp_{12}\wp_{22}+4\wp_{11}+\lambda_3\wp_{22}+\lambda_4\wp_{22}^2+\lambda_2
\end{equation}
one can also find
\begin{equation}
e_ie_j(e_p+e_q+e_r)+e_pe_qe_r=\wp_{11}(\boldsymbol{\Omega}_{i,j}) \,,
\label{JIPE2}
\end{equation}
where $i$,$j$,$p$,$q$, and $r$ are mutually different.

From (\ref{JIPE1}) and (\ref{JIPE2}) we obtain an expression for the matrix $\varkappa$ that is useful for numeric calculations because it reduces the second period matrix to an expression in the first period matrix and $\theta$--derivatives, namely, in the case $e_i=e_1,e_j=e_2$,
\begin{equation}
\varkappa=-\frac12 \left(\begin{array}{cc} e_1e_2(e_3+e_4+e_5)+e_3e_4e_5&-e_1e_2\\
-e_1e_2&e_1+e_2 \end{array}\right)-\frac12 {(2\omega)^{-1}}^T \mathfrak{T}(\boldsymbol{\Omega}_{1,2})(2\omega)^{-1} \,,
\label{kappa2}
\end{equation}
where $\mathfrak{T}$ is the $2\times2$-matrix defined in Proposition \ref{kappaprop}.

\subsection{Inversion of a holomorphic integral}

Taking the limit $z_2\rightarrow \infty$ in the Jacobi inversion problem (\ref{JIP}) we obtain
\begin{equation}
\int_{\infty}^{(z,w)} \frac{\mathrm{d}z}{w}=u_1,\quad \int_{\infty}^{(z,w)} \frac{z\mathrm{d}z}{w}=u_2 \,.
\end{equation}
The same limit in the ratio
\begin{equation}
\frac{\wp_{12}(\boldsymbol{u})}{\wp_{22}(\boldsymbol{u})} = - \frac{z_1z_2}{z_1+z_2}
\end{equation}
leads to the Grant-Jorgenson formula (\ref{gr}).
In terms of $\theta$--functions this can be given the form
\begin{equation}
z = \left.-\frac{\partial_{\boldsymbol{U}}  \theta[\boldsymbol{K}_{\infty}]((2\omega)^{-1}\boldsymbol{u};\tau)}{\partial_{\boldsymbol{V}}  \theta[\boldsymbol{K}_{\infty}]((2\omega)^{-1}\boldsymbol{u};\tau)}\right|_{\theta((2\omega)^{-1}\boldsymbol{u};\tau)=0},
\end{equation}
where here and below $\partial_{\boldsymbol{U}} = \sum_{j=1}^g U_j \frac{\partial}{\partial z_j}$ is the derivative along the direction $\boldsymbol{U}$. Here we introduced the ``winding vectors" $\boldsymbol{U}$, $\boldsymbol{V}$ as column vectors of the inverse matrix
\begin{equation}
(2\omega)^{-1}=(\boldsymbol{U},\boldsymbol{V}) \,.
\end{equation}

From~\eqref{gr} we obtain for all finite branch points
\begin{equation}
e_i=-\frac{\sigma_1(\boldsymbol{\mathfrak A}_i ) }{\sigma_2(\boldsymbol{\mathfrak A}_i) }, \quad i=1,\ldots,5 \label{e2}
\end{equation}
or, equivalently,
\begin{equation}
e_i=-\frac{\partial_{\boldsymbol{U}}\theta[\delta_i]}{\partial_{\boldsymbol{V}}\theta[\delta_i]}, \quad i=1,\ldots,5 \label{e2a} \,.
\end{equation}
This formula was mentioned by Bolza \cite{bolza86} (see his Eq.~(6)) for the case of genus two curve with finite branch points.


\section{Hyperelliptic curve of genus three}\label{sec:curvesgenus3}

As the next case we consider the hyperelliptic curve $X_3$ of genus three with seven real zeros as a model problem in advance to the real physical problems studied in the next section. Let the curve $X_3$ be given by
\begin{align}\begin{split}
w^2&=4(z-e_1)(z-e_2)(z-e_3)(z-e_4)(z-e_5)(z-e_6)(z-e_7)\\
&=4z^{7}+\lambda_6z^6+\ldots+\lambda_1z+\lambda_0 \,.
\end{split}
\label{curve3}
\end{align}

The complete set of holomorphic and meromorphic differentials with a unique pole at infinity is
\begin{align}
\mathrm{d}u_1 & = \frac{\mathrm{d}z}{w}\,, & \qquad \mathrm{d}r_1 & = z(20z^4+4\lambda_6z^3+3\lambda_5z^2+2\lambda_4z+\lambda_3)
 \frac{\mathrm{d}z}{4w} \,, \nonumber \\
\mathrm{d}u_2 &= \frac{z\mathrm{d}z}{w}\,, & \qquad \mathrm{d}r_2 & = z^2(12z^2+2\lambda_6z+\lambda_5)\frac{\mathrm{d}z}{4w}\,, \\
\mathrm{d}u_3 &= \frac{z^2\mathrm{d}z}{w}\,, & \qquad  \mathrm{d}r_3 & = \frac{z^3\mathrm{d}z}{w} \,. \nonumber
\end{align}
Again we introduce the winding vectors
\begin{equation}
(2\omega)^{-1} = \left( \boldsymbol{U}, \boldsymbol{V}, \boldsymbol{W} \right) \, .
\end{equation}

The Jacobi inversion problem for the equations
\begin{align}\begin{split}
\int_{\infty}^{z_1}\frac{ \mathrm{d}z}{w}
+\int_{\infty}^{z_2}\frac{ \mathrm{d}z}{w}
+\int_{\infty}^{z_3}\frac{ \mathrm{d}z}{w}
=u_1,\\
\int_{\infty}^{z_1}\frac{z \mathrm{d}z}{w}
+\int_{\infty}^{z_2}\frac{z \mathrm{d}z}{w}
+\int_{\infty}^{z_3}\frac{z \mathrm{d}z}{w}
=u_2,\\
\int_{\infty}^{z_1}\frac{z^2 \mathrm{d}z}{w}
+\int_{\infty}^{z_2}\frac{z^2 \mathrm{d}z}{w}
+\int_{\infty}^{z_3}\frac{z^2 \mathrm{d}z}{w}
=u_3\end{split} \label{JIP3}
\end{align}
is solved by
\begin{align}\begin{split}
z_1+z_2+z_3&=\wp_{33}(\boldsymbol{u}), \quad z_1z_2+z_1z_3+z_2z_3=-\wp_{23}(\boldsymbol{u}), \quad z_1z_2z_3= \wp_{13}(\boldsymbol{u})\\
w_k&=-\wp_{333}(\boldsymbol{u})z_k^2-\wp_{233}(\boldsymbol{u})z_k -\wp_{133}(\boldsymbol{u}) , \quad k=1,2,3 \,.
\end{split} \label{SOLJIP3}
\end{align}

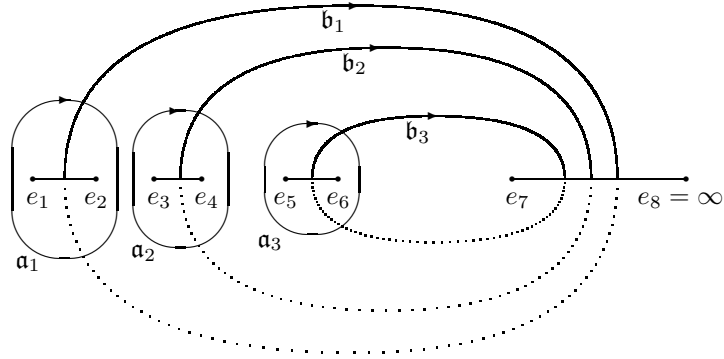
\begin{figure}
\begin{center}
\unitlength 0.7mm \linethickness{0.6pt}
\begin{picture}(150.00,80.00)
\put(9.,33.){\line(1,0){12.}} \put(9.,33.){\circle*{1}}
\put(21.,33.){\circle*{1}} \put(10.,29.){\makebox(0,0)[cc]{$e_1$}}
\put(21.,29.){\makebox(0,0)[cc]{$e_2$}}
\put(15.,33.){\oval(20,30.)}
\put(8.,17.){\makebox(0,0)[cc]{$\mathfrak{ a}_1$}}
\put(15.,48.){\vector(1,0){1.0}}
\put(32.,33.){\line(1,0){9.}} \put(32.,33.){\circle*{1}}
\put(41.,33.){\circle*{1}} \put(33.,29.){\makebox(0,0)[cc]{$e_3$}}
\put(42.,29.){\makebox(0,0)[cc]{$e_4$}}
\put(37.,33.){\oval(18.,26.)}
\put(30.,19.){\makebox(0,0)[cc]{$\mathfrak{a}_2$}}
\put(36.,46.){\vector(1,0){1.0}}
\put(57.,33.){\line(1,0){10.}} \put(57.,33.){\circle*{1}}
\put(67.,33.){\circle*{1}} \put(57.,29.){\makebox(0,0)[cc]{$e_5$}}
\put(67.,29.){\makebox(0,0)[cc]{$e_6$}}
\put(62.,33.){\oval(18.,21.)}
\put(54.,21.){\makebox(0,0)[cc]{$\mathfrak{a}_3$}}
\put(62.,43.5){\vector(1,0){1.0}}
\put(100.,33.00) {\line(1,0){33.}} \put(100.,33.){\circle*{1}}
\put(133.,33.){\circle*{1}}
\put(101.,29.){\makebox(0,0)[cc]{$e_{7}$}}
\put(132.,29.){\makebox(0,0)[cc]{$e_{8}=\infty$}}
\put(66.,63.){\makebox(0,0)[cc]{$\mathfrak{b}_1$}}
\put(70.,66.){\vector(1,0){1.0}}
\bezier{484}(15.,33.)(15.,66.)(70.,66.)
\bezier{816}(70.,66.)(120.,66.00)(120.,33.)
\bezier{35}(15.,33.)(15.,0.)(70.,0.)
\bezier{35}(70.,0.)(120.,0.)(120.,33.)
\put(70.,55.){\makebox(0,0)[cc]{$\mathfrak{b}_2$}}
\put(74.00,58.){\vector(1,0){1.0}}
\bezier{384}(37.,33.)(37.,58.)(76.,58.)
\bezier{516}(76.,58.)(115.,58.)(115.00,33.00)
\bezier{30}(37.,33.00)(37.,8.)(76.00,8.)
\bezier{30}(76.00,8.)(115.00,8.00)(115.00,33.00)
\put(82.,42.){\makebox(0,0)[cc]{$\mathfrak{b}_3$}}
\put(85,45.){\vector(1,0){1.0}}
\bezier{384}(62.,33.)(62.,45.)(85.,45.)
\bezier{516}(85.,45.)(110.,45.)(110.00,33.00)
\bezier{30}(62.,33.00)(62.,21.)(85.00,21.)
\bezier{30}(85.00,21.)(110.00,21.00)(110.00,33.00)
\end{picture}
\end{center}
\caption{Homology basis on the Riemann surface of the curve
$X_3$ with real branch points $e_1 < e_2 <\ldots <
e_{8}=\infty$ (upper sheet).  The cuts are drawn from $e_{2i-1}$
to $e_{2i}$, $i=1,2,4$.  The $\mathfrak{b}$--cycles are completed on the
lower sheet (dotted lines).} \label{figure-3}
\end{figure}

\subsection{Characteristics in genus three}

Let $\mathfrak{A}_k$ be the Abelian image of the $k$-th branch point, namely
\begin{equation}
\boldsymbol{\mathfrak{A}}_k=\int_{\infty}^{e_k} \mathrm{d}\boldsymbol{u}= 2\omega \boldsymbol{\varepsilon}_k+2\omega' \boldsymbol{\varepsilon}_k', \quad k=1,\ldots,8 \,,
\end{equation}
where $\boldsymbol{\varepsilon}_k$ and $\boldsymbol{\varepsilon}_k'$ are column vectors whose entries $\varepsilon_{k,j}$, $\varepsilon'_{k,j}$ are $\frac{1}{2}$ or zero for all $k=1,\ldots,8$, $j=1,2,3$.

The correspondence between branch points and characteristics in the fixed homology basis is given as
\begin{align}\begin{split}
[\boldsymbol{{\mathfrak A}}_1]= \frac{1}{2}
\begin{pmatrix} 1 & 0 & 0 \\ 0 & 0 & 0 \end{pmatrix}\, ,\quad
[\boldsymbol{{\mathfrak A}}_2]= \frac{1}{2}
\begin{pmatrix} 1 & 0 & 0 \\ 1 & 0 & 0 \end{pmatrix}\, ,\quad
[\boldsymbol{{\mathfrak A}}_3] = \frac{1}{2}
\begin{pmatrix} 0 & 1 & 0 \\ 1 & 0 & 0 \end{pmatrix}\, , \\
[\boldsymbol{{\mathfrak A}}_4]= \frac{1}{2}
\begin{pmatrix} 0 & 1 & 0 \\ 1 & 1 & 0 \end{pmatrix}\, ,\quad
[\boldsymbol{{\mathfrak A}}_5] = \frac{1}{2}
\begin{pmatrix} 0 & 0 & 1 \\ 1 & 1 & 0 \end{pmatrix}\, ,\quad
[\boldsymbol{{\mathfrak A}}_6]= \frac{1}{2}
\begin{pmatrix} 0 & 0 & 1 \\ 1 & 1 & 1 \end{pmatrix}\, , \\
[\boldsymbol{{\mathfrak A}}_7]= \frac{1}{2}
\begin{pmatrix} 0 & 0 & 0 \\ 1 & 1 & 1 \end{pmatrix}\, ,\quad
[\boldsymbol{{\mathfrak A}}_8]= \frac{1}{2}
\begin{pmatrix} 0 & 0 & 0 \\ 0 & 0 & 0 \end{pmatrix}\, . \end{split}
\label{hombasis_gen3}
\end{align}

The vector of Riemann constants $\boldsymbol{K}_{\infty}$ with the base point at infinity is given in the above basis by the even singular characteristics,
\begin{equation}
[\boldsymbol{K}_{\infty}]=[\boldsymbol{\mathfrak A}_2]+[\boldsymbol{\mathfrak A}_4]+[\boldsymbol{\mathfrak A}_6] = \frac12 \begin{pmatrix} 1 & 1 & 1 \\ 1 & 0 & 1 \end{pmatrix}  \, .
\end{equation}

From the above characteristics 64 half-periods can be built as follows. Start with singular even characteristics, there should be only one such characteristic that corresponds to the vector of Riemann constants $\boldsymbol{K}_{\infty}$. The corresponding partition reads $\mathcal{I}_2\cup \mathcal{J}_2 = \{ \} \cup \{1, 2, \ldots, 8\}$ and the $\theta$--function $\theta(\boldsymbol{K}_{\infty}+\boldsymbol{v})$ vanishes at the origin $\boldsymbol{v}=0$ to the order $m=2$.

The half-periods $\boldsymbol{\Delta}_1=(2\omega)^{-1}\boldsymbol{\mathfrak A}_k+\boldsymbol{K}_{\infty}\in \Theta_1$ correspond to partitions
\begin{equation}
\mathcal{I}_1\cup \mathcal{J}_1 =  \{ k,8 \} \cup \{ j_1,\ldots, j_4 \}, \quad j_1,\ldots,j_4 \notin \{8,k\}
\end{equation}
and the $\theta$--function $\theta(\boldsymbol{\Delta}_1+\boldsymbol{v})$ vanishes at the origin $\boldsymbol{v}=0$ to the order $m=1$.

Also denote the $21$ half-periods that are images of two branch points
\begin{align}
\boldsymbol{\Omega}_{i,j}=2\omega ( \boldsymbol{\varepsilon}_i+\boldsymbol{\varepsilon}_j)+2\omega' (\boldsymbol{\varepsilon}_i'+  \boldsymbol{\varepsilon}_j') ,\quad  i,j = 1,\ldots, 7, i\neq j \ .
\label{odd2}
\end{align}
The half-periods $\widehat{\boldsymbol{\Delta}}_1=(2\omega)^{-1}\boldsymbol{\Omega}_{i,j}
+\boldsymbol{K}_{\infty}\in \Theta_2$ correspond to the partitions
\begin{equation}
\mathcal{I}_1\cup \mathcal{J}_1 = \{ i,j \} \cup \{ j_1,\ldots, j_4 \}, \quad j_1,\ldots,j_4 \notin \{i,j\}
\end{equation}
and the $\theta$--function $\theta(\widehat{\boldsymbol{\Delta}}_1 + \boldsymbol{v})$ vanishes at the origin, $\boldsymbol{v}=0$, as before to the order $m=1$.
Therefore the characteristics of the $7$ half-periods
\begin{equation}
\left[(2\omega)^{-1} \boldsymbol{\mathfrak A}_i+\boldsymbol{K}_{\infty}\right] =: \delta_i \, ,\quad i=1,\ldots,7
\end{equation}
are nonsingular and odd as well as the characteristics of the $21$ half-periods
\begin{equation}
\left[(2\omega)^{-1} \boldsymbol{\Omega}_{i,j}+\boldsymbol{K}_{\infty}\right] =: \delta_{i,j}   \, ,\quad 1\leq i<j\leq 7 \,.
\end{equation}
We finally introduce the $35$ half-periods that are images of three branch points
\begin{align}
\boldsymbol{\Omega}_{i,j,k}=2\omega ( \boldsymbol{\varepsilon}_i+\boldsymbol{\varepsilon}_j+\boldsymbol{\varepsilon}_k)+ 2\omega' (\boldsymbol{\varepsilon}_i'+  \boldsymbol{\varepsilon}_j'+  \boldsymbol{\varepsilon}_k')\in \mathrm{Jac}(X_g) ,\quad  1\leq i<j<k\leq 7 \,.
\label{even3}
\end{align}
The half-periods $\widehat{\widehat{\boldsymbol{\Delta}}}_1=(2\omega)^{-1}\boldsymbol{\Omega}_{i,j,k}
+\boldsymbol{K}_{\infty}$ correspond to the partitions
\begin{equation}
\mathcal{I}_0\cup \mathcal{J}_0 =  \{ i,j,k,8 \} \cup \{ j_1,\ldots, j_4 \}, \quad j_1,\ldots,j_4 \notin \{i,j,k,8\}\,.
\end{equation}
The $\theta$--function $\theta(\widehat{\widehat{\boldsymbol{\Delta}}}_1+\boldsymbol{v})$
does not vanish at the origin $\boldsymbol{v}=0$.

Furthermore, the $35$ characteristics
\begin{equation}
[\varepsilon_{i,j,k}]=  \left[ (2\omega)^{-1} \boldsymbol{\Omega}_{i,j,k}+\boldsymbol{K}_{\infty} \right],\quad 1\leq i<j<k\leq 7
\end{equation}
are even and nonsingular while the characteristic $[\boldsymbol{K}_{\infty}]$ is even and singular. Altogether we got all $64=4^3$ characteristics classified by the partitions of the branch points.

\subsection{Inversion of a holomorphic integral}

All three holomorphic integrals,
\begin{align}
\int_{\infty}^x \frac{\mathrm{d}z}{w}=u_1,\quad
\int_{\infty}^x \frac{z\mathrm{d}z}{w}=u_2,\quad
\int_{\infty}^x \frac{z^2\mathrm{d}z}{w}=u_3
\end{align}
are inverted by the same formula (\ref{onishi}). Nevertheless, there are three different cases for which one of the variables $u_1,u_2,u_3$ is considered as independent while the remaining two result from solving the divisor conditions $\sigma(\boldsymbol{u})=\sigma_3(\boldsymbol{u})=0$.

Formula (\ref{onishi}) can be rewritten in terms of $\theta$--functions as
\begin{align}
x=-\frac{\partial^2_{\boldsymbol{U},\boldsymbol{W}}
\theta[\boldsymbol{K}_{\infty}]((2\omega)^{-1}\boldsymbol{u})
+2(\partial_{\boldsymbol{U}}
\theta[\boldsymbol{K}_{\infty}]((2\omega)^{-1}\boldsymbol{u}))
\boldsymbol{e}_3^T\varkappa\boldsymbol{u}}{\partial^2_{\boldsymbol{V},\boldsymbol{W}}
\theta[\boldsymbol{K}_{\infty}]((2\omega)^{-1}\boldsymbol{u})
+2(\partial_{\boldsymbol{V}}
\theta[\boldsymbol{K}_{\infty}]((2\omega)^{-1}\boldsymbol{u}))
\boldsymbol{e}_3^T\varkappa\boldsymbol{u}}, \label{xtheta}
\end{align}
where $\boldsymbol{e}_3=(0,0,1)^T$. This represents the solution of the inversion problem.

From the solution of the Jacobi inversion problem follows for any $1\leq i<j<k \leq 7$,
\begin{equation}
e_i+e_j+e_k=\wp_{33}(\boldsymbol{\Omega}_{i,j,k}),\quad
-e_ie_j-e_ie_k-e_je_k=\wp_{23}(\boldsymbol{\Omega}_{i,j,k}),\quad
e_ie_je_k=\wp_{13}(\boldsymbol{\Omega}_{i,j,k}) \,.
\label{JIPE31}
\end{equation}
From (\ref{cubic}) we get the relations
\begin{align}
\wp_{333}^2 &= 4\wp_{33}^3+\lambda_6\wp_{33}^2+4\wp_{23}\wp_{33}+\lambda_5\wp_{33}+4\wp_{22}-4\wp_{13}+\lambda_4 \,, \nonumber \\
\wp_{233}^2 &= 4\wp_{23}^2\wp_{33}+\lambda_6\wp_{23}^2-4\wp_{22}\wp_{23}+8\wp_{13}\wp_{23}+4\wp_{11}+\lambda_2 \,,\\
\wp_{133}^2 &= 4\wp_{13}^2\wp_{33}+\lambda_6\wp_{13}^2-4\wp_{12}\wp_{13}+\lambda_0 \nonumber
\end{align}
and also find
\begin{equation}
\wp_{12}(\boldsymbol{\Omega}_{i,j,k})=-s_3S_1-S_4 \, , \quad \wp_{11}(\boldsymbol{\Omega}_{i,j,k})=s_3S_2+s_1S_4 \, , \quad \wp_{22}(\boldsymbol{\Omega}_{i,j,k})=S_3+2s_3+s_2S_1 \,,
\label{JIPE32}
\end{equation}
where $s_l$ are the elementary symmetric functions of order $l$ of the branch points $e_i,e_j,e_k$ and $S_l$ are the elementary symmetric functions of order $l$ of the remaining branch points $\{1,\ldots,7\} \setminus \{i,j,k\}$.

From (\ref{JIPE31}) and (\ref{JIPE32}) one can find the following expression for the matrix $\varkappa$
\begin{align}
\varkappa = -\frac12 \mathfrak{P}((\boldsymbol{\Omega}_{i,j,k})) -\frac12 {(2\omega)^{-1}}^T \mathfrak{T}(\boldsymbol{\Omega}_{i,j,k}) (2\omega)^{-1}
\label{kappa3}
\end{align}
where $i,j,k$ are arbitrary and $\mathfrak{T}(\boldsymbol{\Omega}_{i,j,k})$ is the $3\times3$-matrix defined in Proposition \ref{kappaprop},
\begin{equation}
\mathfrak{T}(\boldsymbol{\Omega}_{i,j,k}) = \left(
-\frac{\partial^2}{\partial z_m \partial z_n  }\mathrm{log}\, \theta[\boldsymbol{K}_{\infty}]((2\omega)^{-1}\boldsymbol{\Omega}_{i,j,k};\tau )
\right)_{m,n=1,2,3} \, .
\end{equation}

For the branch points $e_1,\ldots, e_8$ the expression
\begin{equation}
e_i=-\frac{\partial_{\boldsymbol{U}}\left[
\partial_{\boldsymbol{W}}+2\boldsymbol{\mathfrak A}^T_i\varkappa \boldsymbol{e}_3
\right]\theta[\boldsymbol{K}_{\infty}]((2\omega)^{-1} \boldsymbol{\mathfrak A}_i;\tau)}
{\partial_{\boldsymbol{V}}\left[
\partial_{\boldsymbol{W}}+2\boldsymbol{\mathfrak A}^T_i\varkappa \boldsymbol{e}_3
\right]\theta[\boldsymbol{K}_{\infty}]((2\omega)^{-1} \boldsymbol{\mathfrak A}_i;\tau)}
\label{thomae1}
\end{equation}
is valid. Furthermore we have for $i,j=1,\ldots,8$, $i\neq j$
\begin{align}\begin{split}
e_i + e_j & = - \frac{\sigma_2(\boldsymbol{\Omega}_{i,j})}
{\sigma_3(\boldsymbol{\Omega}_{i,j})} \equiv 
-\frac{\partial_{\boldsymbol{V}} \theta[\delta_{i,j}]}
{\partial_{\boldsymbol{W}} \theta[\delta_{i,j}]}, \\
e_i e_j & = \frac{\sigma_1(\boldsymbol{\Omega}_{i,j})}
{\sigma_3(\boldsymbol{\Omega}_{i,j})} \equiv 
\frac{\partial_{\boldsymbol{U}} \theta[\delta_{i,j}]}
{\partial_{\boldsymbol{W}} \theta[\delta_{i,j}]} \, ,
\end{split}\label{thomae2}
\end{align}
and for $i=1,\ldots,7$
\begin{equation}
e_i = -\frac{\sigma_1(\boldsymbol{\mathfrak A}_i)} {\sigma_2(\boldsymbol{\mathfrak A}_i)} 
= - \frac{\partial_{\boldsymbol{U}} , \theta[\delta_{i}]}
{\partial_{\boldsymbol{V}} \theta[\delta_{i}]} \, .
\end{equation}

Using these data one can reconstruct by the known matrices $2\omega,2\omega'$ all the characteristics $[\boldsymbol{\mathfrak{A}}_j]$, $j=1,\ldots, 2g+2$, as well as the vector of Riemann constants following the procedure given in the Sec.~\ref{moduli_calc}. For example, for the genus 3 curve
\begin{equation}
y^2=4x(x-1)(x-2)(x-3)(x-4)(x-5)(x-6)
\end{equation}
we construct the ``Maple/alcurves" homology basis and compute the Riemann period matrices $(2\omega,2\omega')$. We order the branch points according to
\begin{equation}
e_1=0,\;\; e_2=1,\;\; e_3=2,\;\; e_4=3,\;\;e_5=4,\;\; e_6=5,\;\;e_7=6,\;\;e_8 = \infty \, .
\end{equation}
Then we find the Abelian images of the branch points
\begin{align}
[\boldsymbol{\mathfrak{A}}_1] & = \frac12 \begin{pmatrix} 1 & 0 & 0 \\ 1 & 1 & 1 \end{pmatrix} , & \quad
[\boldsymbol{\mathfrak{A}}_2] & = \frac12 \begin{pmatrix} 1 & 0 & 0 \\ 0 & 1 & 1 \end{pmatrix} , & \quad
[\boldsymbol{\mathfrak{A}}_3] & = \frac12 \begin{pmatrix} 0 & 1 & 0 \\ 0 & 0 & 1 \end{pmatrix} , \nonumber \\
[\boldsymbol{\mathfrak{A}}_4] & = \frac12 \begin{pmatrix} 0 & 0 & 1 \\ 0 & 0 & 0 \end{pmatrix} , & \quad
[\boldsymbol{\mathfrak{A}}_5] & = \frac12 \begin{pmatrix} 0 & 1 & 0 \\ 0 & 1 & 1 \end{pmatrix} , & \quad
\boldsymbol{\mathfrak{A}}_6] & = \frac12 \begin{pmatrix} 0 & 0 & 1 \\ 0 & 0 & 1 \end{pmatrix} , \\
[\boldsymbol{\mathfrak{A}}_7] & = \frac12 \begin{pmatrix} 0 & 0 & 0 \\ 1 & 1 & 1 \end{pmatrix} , & \quad
[\boldsymbol{\mathfrak{A}}_8] & = \frac12 \begin{pmatrix} 0 & 0 & 0 \\ 0 & 0 & 0 \end{pmatrix} \, . & & \nonumber
\end{align}
There are 3 odd among these characteristics and therefore
\begin{equation}
[\boldsymbol{K}_{\infty}] = \frac12 \begin{pmatrix} 1 & 1 & 1 \\ 1 & 0 & 1 \end{pmatrix} \, .
\end{equation}
One can check that $[\boldsymbol{K}_{\infty}]$ is the only even characteristic such that $\theta([\boldsymbol{K}_{\infty}])=0$ as follows from Table~\ref{table1}.
Now one can construct arbitrary nonsingular even characteristics and calculate $\varkappa$ and then the second period matrix.

\section{An application: $9$-dimensional Reissner-Nordstr\"om-de Sitter space-time}\label{sec:application}

Being armed with the method developed we now come back to one of the physical model systems presented in Section~\ref{sec:particlemotion}. We consider here the geodesic equation in a Reissner-Norstr\"om-de Sitter space-time of nine dimension. Setting $d=9$ and introducing a new dimensionless coordinate $\tr = r/r_{\rm S}$ and the dimensionless parameters
$\tl = L/r_{\rm S}$, $\tq = q/r_{\rm S}$, and $\tla = \Lambda r^2_{\rm S}$
we obtain from \eqref{beglkusy}
\begin{equation}
\left(\frac{d\tr}{d\varphi}\right)^2 = \frac{R_{16}(\tr)}{\tr^{10}} \, ,  \label{beglkusy2} \end{equation}
where
\begin{equation}
\tl^2 R_{16}(\tr) = \tr^{16} \delta\frac{\tla}{28} + \left(E^2 - \delta + \frac{\tla}{28} \tl^2\right) \tr^{14} - \tr^{12} \tl^2 + \tr^8 \delta + \tr^6 \tl^2 - \tq^2 \delta \tr^2 - \tq^2 \tl^2 \, .
\end{equation}
Through the substitution $\tr= 1/\sqrt{u}$ we halve the order of the polynomial and by  $u = x^{-1} + u_8$, where $u_8$ is any root of the polynomial $R_8(u)$, we obtain a polynomial of seventh order so that Eq.~\eqref{beglkusy2} reduces to
\begin{equation}
\left(x^2\frac{dx}{d\varphi}\right)^2 = \mathcal{P}_7(x) \equiv \sum^{7}_{i=0}b_ix^i=b_7\prod^7_{i=1}(x-e_i) \, .
\end{equation}
Thus, solving the differential equation is reduced to the inversion of a genus three holomorphic hyperelliptic integral
\begin{equation}
\varphi-\varphi_{\rm in}=\int^{x}_{x_{\rm in}} \frac{x^{\prime^2}dx^\prime}{\sqrt{\mathcal{P}_7(x^\prime)}} \ ,
\end{equation}
where $\varphi_{\rm in}$ denotes the initial conditions for $\varphi$ and $x_{\rm in}$ is the starting point of the integration.

From \eqref{onishi} we find the solution of the equation of motion~\eqref{beglkusy2}
\begin{equation}
\tr(\varphi) = \frac{1}{\sqrt{- \dfrac{\sigma_{23}(\boldsymbol{u})}{\sigma_{13}(\boldsymbol{u})}  + u_8 } } \ ,
\end{equation}
where \begin{equation}  \boldsymbol{u}=\boldsymbol{\mathfrak A}_i+\left( \begin{array}{c}  f_1(\varphi - \varphi_{\rm in}) \\ f_{2}(\varphi - \varphi_{\rm in}) \\ \varphi - \varphi_{\rm in} \end{array}   \right) ,\quad f_1(0)=f_2(0)=0 \ , \label{u1u2u3}
\end{equation}
and where the functions $f_1(\varphi - \varphi_{\rm in})$ and $f_2(\varphi - \varphi_{\rm in})$ can be found from the conditions $\sigma(\boldsymbol{u})=0$ and $\sigma_3(\boldsymbol{u})=0$. Also $x_{\rm in}$ is chosen as a branch point of the polynomial $\mathcal{P}_7(x)$ which defines the half-integer characteristic $\boldsymbol{\mathfrak A}_i$. The homology basis is fixed by the characteristics~\eqref{hombasis_gen3} and the $\sigma$-quotient is computed according to the
formula (\ref{xtheta}) while the genus three $\theta$--functions were calculated with the aid of the ''Maple/RiemannTheta`` code.

\begin{figure}[t]
\begin{center}
\subfigure[][$E^2$-$\tl^2$-diagram: the dark gray region indicates many--world periodic bound and escape orbits, as e.g. the orbits~\subref{sds9d_b} and~\subref{sds9d_e}. In the light gray region there is one two-world escape orbit.]{\label{diagram}\includegraphics[width=0.3\textwidth]{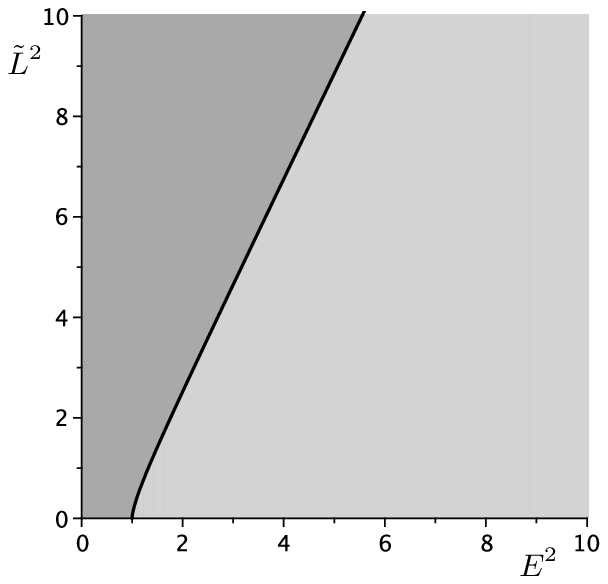}}\qquad
\subfigure[][many--world periodic bound orbit (for a review of the classification of the geodesics in Reissner--Nordstr\"om--(anti) de Sitter space--times see~\cite{Hackmannetal08}.)]{\label{sds9d_b}\includegraphics[width=0.3\textwidth]{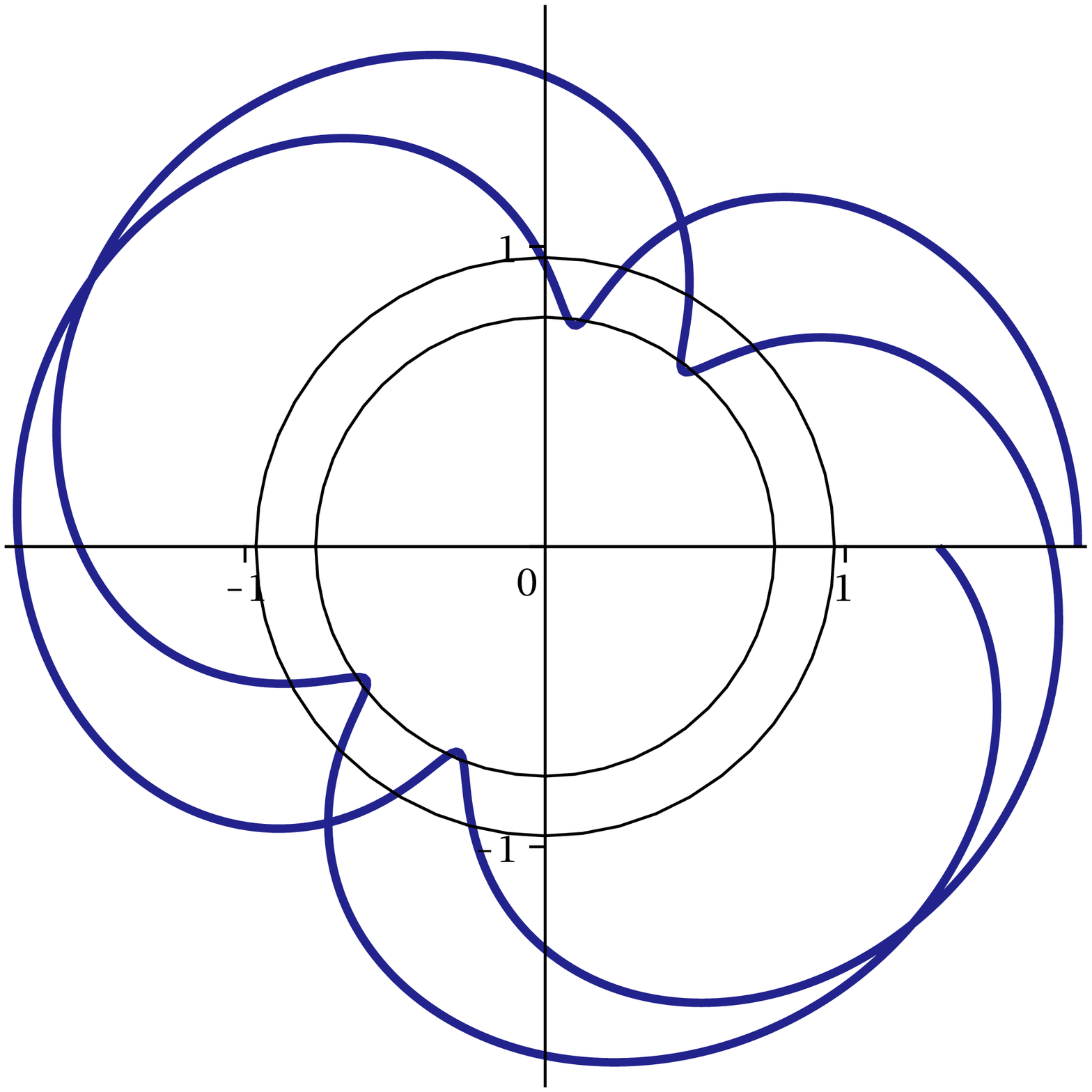}}\qquad
\subfigure[][escape orbit]{\label{sds9d_e}\includegraphics[width=0.3\textwidth]{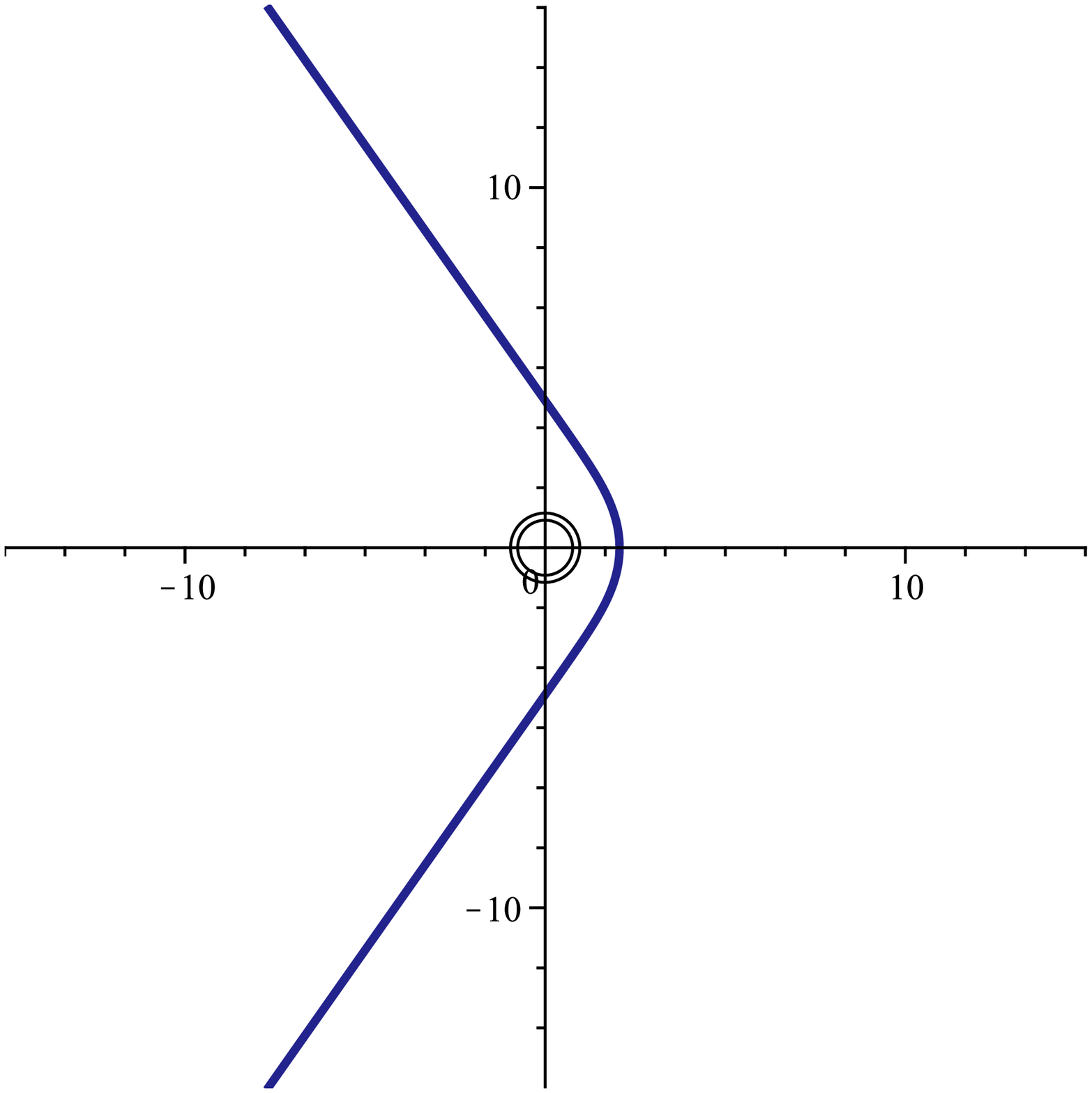}}
\end{center}
\caption{Examples of test particle motion in a 9--dimensional Reissner--Nordstr\"om--de Sitter space--time. The chosen parameters for the cosmological constant and the charge are $\tla=8.7 \cdot 10^{-5}$, $\tq=0.4$. We also consider massive particles, $\delta = 1$. For these parameters \subref{diagram} shows the $E^2-\tl^2$ parameter plot where the dark area indicates three positive real zeros and the bright area 1 real positive zero of $\mathcal{P}_7$. For the chosen values $E^2=1.045$ and $\tl=0.5$ two orbits are calculated analytically and plotted in  \subref{sds9d_b} and \subref{sds9d_e}. The black circles are the event and the Cauchy horizons. In this example $e_1<e_2<e_3<e_4<\mathrm{Re}(e_5)<e_7$, and $e_5=\overline{e_6}$. For the many-world periodic bound orbit~\subref{sds9d_b} we choose a starting point $e_2$, which corresponds to the half period $[{\mathfrak A}_2]$ in~\eqref{hombasis_gen3}, and the motion is bounded by $e_1$ and $e_2$. For the escape orbit~\subref{sds9d_e} the starting point is $e_3$, which corresponds to the half period $[{\mathfrak A}_3]$ in~\eqref{hombasis_gen3}, and the test particle moves to infinity. \label{sds9d}}
\end{figure}

The structure of the orbits is given by the number of zeros of the polynomial $\mathcal{P}_7$. This depends on the choice of the parameters $E$, $\tilde L$, $\tilde\Lambda$, and $\tilde q$. We choose a certain nontrivial value for $\tilde\Lambda$ and $\tilde q$ and draw the $E^2 - \tilde L^2$ parameter plot presented in Fig.~\ref{sds9d}\subref{diagram}. In the dark area the polynomial $\mathcal{P}_7$ possesses 3 positive zeros resulting in one bound and one escape orbit, and in the brighter area it possesses one positive zero, yielding one two--world escape orbit. For certain values for $E^2$ and $\tl^2$ we obtain analytically the orbits shown in Fig.~\ref{sds9d}. From the causal structure of the metric encoded in the zeros of $g_{tt} = g_{rr}^{-1}$ one concludes that after traversing both horizons a second time the test body enters a new universe (see, e.g., the discussion in \cite{Hackmannetal08}).

Furthermore, the observable perihelion shift $\Omega_{\rm perihelion}$ can be given by a complete hyperelliptic integral
\begin{equation}
\Omega_{\rm perihelion} = 2 \int_{r_{\rm min}}^{r_{\rm max}} \frac{y^2}{\sqrt{\mathcal{P}_7(y)}} dy - 2 \pi \, .
\end{equation}
The perihelion shift compares the period of the radial motion with $2\pi$.

\section{Conclusion}

In this paper we presented the analytical solution of geodesic equations in general relativistic models, containing a hyperelliptic curve of arbitrary genus. Based on the solution for genus $2$ and genus $3$ and the investigation of the Schur--Weierstra{\ss} polynomials, we present the solution~\eqref{matprev} for the geodesic equations with an underlying curve of arbitrary genus which is of the form or can be reduced to the form
\begin{equation}
\varphi - \varphi_{\rm in} = \int^x_{x_{\rm in}} y^i \frac{dy}{\sqrt{\mathcal{P}_{2g+1}(y)}} \, , \qquad i=0, \ldots, g-1 \, ,
\end{equation}
where $g$ is the genus of the curve $w^2=\mathcal{P}_{2g+1}(x)$. As an example we integrate the geodesic equations with an underlying hyperelliptic curve of genus $3$ in the Reissner--Nordstr\"om--de Sitter space--time in $9$ dimensions.

One major task in the calculation of the analytical solution is the calculation of the first and second period matrices. In the Proposition~\ref{kappaprop} we provide a convenient and quick method for the calculation of the matrix $\varkappa$ and the second period matrix from the first period matrix. The matrix $\varkappa$ appears in the $\sigma$--function and, thus, in the general solution~\eqref{matprev}. In this method no integration of meromorphic differentials is required and the result is given in terms of the holomorphic period matrix and $\theta$--constants.

In Section~\ref{moduli_calc} we also propose a step by step algorithm for the calculation of the characteristics of the branch points and the vector of Riemann constants which appear in the $\theta$-- and $\sigma$--functions and which are required for the calculation of the matrix $\varkappa$. We note that all the calculations can be done with the ``Maple/algcurves" package, working in the homology basis automatically chosen by the computer program.

The proposed solution~\eqref{matprev} for the curves of $g\geq 3$ together with the solution for genus $2$ curves has a wide spectrum of applications. These range from the geodesic equations in a wide class of black holes space--times like Schwarzschild--de Sitter~\cite{HackmannLaemmerzahl08,HackmannLaemmerzahl08a}, Kerr--de Sitter~\cite{Hackmannetal010}, NUT--de Sitter~\cite{NdSprep} and general type D Pleba\'{n}ski--Demia\'{n}ski~\cite{Hackmannetal09} space-times in $4$ dimensions to the higher dimensional spherically symmetric Schwarzschild--(anti-)de Sitter, Reissner--Nordstr\"om--(anti-)de Sitter space--times~\cite{Hackmannetal08}, higher dimensional axially symmetric Myers--Perry space--times~\cite{MPprep} discussed in Section~\ref{sec:particlemotion}, and general Kerr--NUT--(anti-)de Sitter metrics in all dimensions introduced in~\cite{ChLP06}. This might be extended to space--times with cosmic strings \cite{Hackmannetal10}, even in higher dimensions. The mathematical methods described here are also applicable to such problems as the motion of test particles with spin \cite{HackmannLaemmerzahlSchaffer10} and mass multipoles and to the motion of test particles with and without spin and mass multipoles in gravitating mass multipole fields, which have applications in astrophysics, satellite dynamics, and geodesy. The motion of test particles in the gravitational field of a black hole with disturbances related to mass multipoles have been discussed as test of the no--hair theorem \cite{Will09}. As a consequence, there is a wide range of applications of the formalism developed in the article to problems in the area of General Relativity.





\section*{Acknowledgments} The authors would like to thank H.~Braden, Y.~Fedorov, B.~Hartmann, 
Sh.~Matsutani, Yo.~$\widehat{\rm O}$nishi, E.~Previato, and P.~Richter for fruitful discussions. E.H. and V.K. acknowledge the financial support of the German Research Foundation DFG, and V.E. the financial support from the  Hanse--Wissenschaftskolleg (Institute for Advanced Study) in Delmenhorst as well as its hospitality. C.L. thanks the center of excellence QUEST for support.

\providecommand{\bysame}{\leavevmode\hbox to3em{\hrulefill}\thinspace}


\begin{thebibliography}{EPR02}

\bibitem[AF00]{abendfed00}
S.~Abenda and Yu.~Fedorov, On the weak Kowalevski-Painlev\'e property for hyperelliptically separable systems, {\em Acta Appl.Math}. \textbf{60} (2000), 137.

\bibitem[BE55] {be55}
H.~Bateman and A.~Erdelyi, \emph{Higher {T}ranscendental {F}unctions}, {\bf{2}}
McGraw-Hill, New York, 1955.

\bibitem[Bak897]{ba97}
H.~F.~Baker, \emph{Abel's theorem and the allied theory of theta functions},
  Cambridge Univ. Press, Cambridge, 1897, Reprinted in 1995.

\bibitem[Bak907]{ba907}
H.~F.~Baker, \emph{Multiple periodic functions} (Cambridge Univ. Press, Cambridge, 1907).

\bibitem[Bak903]{ba03}
H.~F.~Baker, \emph{On a system of differential equations leading to periodic functions}, Acta Math. \textbf{27} (1903), 135.

\bibitem[BG06]{bg06}
S.~Baldwin and J.~Gibbons, \emph{Genus 4 trigonal reduction of the Benney equations}, J. Phys. A  39  (2006) 3607.

\bibitem[BG04]{bg04}
S.~Baldwin and J.~Gibbons, \emph{Higher genus hyperelliptic reductions of the Benney equations}, J. Phys. A  37  (2004) 5341.

\bibitem[Bol886]{bolza86} O.~Bolza,
\emph{Ueber die {R}eduction hyperelliptischer {I}ntegrale erster {O}rdnung und erster {G}attung auf elliptische durch eine {T}ransformation vierten {G}rades}, Math. Ann., \textbf{XXVIII} (1886), 447.

\bibitem[BEH05]{beh05} H.~W.~Braden and V.~Z.~Enolskii and A.~N.~W.~Hone
\emph{Bilinear recurrences and addition formulae for hyperelliptic sigma functions},
J. Nonlinear Math. Phys.  {\bf 12}  (2005),  suppl. 2, 46.

\bibitem[BEL97]{BEL97}
V.~M.~Buchstaber and V.~Z.~Enolskii and D.~V.~Leykin, \emph{Kleinian functions, hyperelliptic Jacobians and applications.}, Reviews in Mathematics and Math. Physics, I.~M.~Krichever, S.~P.~ Novikov Editors, Vol.~10, part~2 (Gordon and Breach, London, 1997), p.3.

\bibitem[BEL99]{BEL99} V.~M.~Buchstaber and V.~Z.~Enolskii and D.~V.~Leykin,
\emph{Rational analogues of Abelian functions}, Funk. Anal. Appl., \textbf{33} (1999), 1.

\bibitem[BL05]{BL05} V.~M.~Buchstaber and and D.~V.~Leykin,
\emph{Addition laws on {J}acobian variety of plane  algebraic curves},
Proceedings of the Steklov Institute of Mathematics, \textbf{251} (2005) 1.

\bibitem[BD99]{BuonannoDamour99}
A.~Buonanno and T.~Damour, \emph{Effective one-body approach to general relativistic two-body dynamics}, Phys. Rev. {\bf D 59} (1999) 084006.

\bibitem[ChLP06]{ChLP06}
W.~Chen, H.~L\"u, C. N.~Pope, \emph{General Kerr-NUT-AdS metrics in all dimensions}, Class. Quantum Grav. \textbf{23} (2006), 5323.

\bibitem[DJS00]{DamourJaranowskiSchaefer00a}
T.~Damour, P. Jaranowski, and G. Sch{\"a}fer, \emph{On the determination of the last stable orbit for circular general relativistic binaries at the third post-Newtonian approximation}, {Phys. Rev.} {\bf D 62} (2000) 084011.

\bibitem[EEKL93]{eekl93}
J.~C.~Eilbeck and  V.~Z.~Enolskii and A.~B.~Kuznetsov and D.~V.~Leykin, {\emph Linear $r$r-matrix algebra for systems separable in parabolic coordinates}, Phys. Lett. {\bf A 180} (1993) 208.

\bibitem[Emp08]{Emparan2008}
R.~Emparan, \emph{Black holes galore in $D>4$}, Fortschr. Phys. {\bf 7-9} (2008), 723.

\bibitem[EK11]{enko11}
V.~Z. Enolski, Ya.~Kopeliovich, \emph{On the generalized Thomae formulae},
in preparation.

\bibitem[EPR03]{EPR03}
V.~Z. Enolskii, M.~Pronine, and P.~Richter, \emph{Double pendulum and theta-divisor}, J. Nonlin. Sci. {\bf 13} (2003) 157.

\bibitem[FK80]
{fk80}H.~ M.~Farkas and I.~Kra, \emph{Riemann {S}urfaces} (Springer, New York, 1980).

\bibitem[Fay73]{fa73}
J.~D.~Fay, \emph{Theta functions on {R}iemann surfaces}, Lectures Notes in
 Mathematics, {\bf 352}, Springer, Berlin 1973.

\bibitem[FG07]{fg07}
Yu.~N.~Fedorov and D.~G\'omes-Ulate \emph{Dynamical systems on infinitely sheeted {Riemann} surfaces}, Physica D {\bf 227} 120.

\bibitem[Gra90]{gr90}
D.~Grant, \emph{Formal groups in genus two}, J. reine angew. Math. \textbf{411} (1990) 96-121

\bibitem[KKHL10]{KKHL10}
V.~Kagramanova, J.~Kunz, E.~Hackmann and C.~L\"ammerzahl, \emph{Analytic treatment of complete and incomplete geodesics in Taub--NUT space--times}, Phys. Rev. D {\bf 81} (2010) 124044.

\bibitem[G02]{Goldstein02}
H.~Goldstein, C.~Poole, and J.~Safko, \emph{Classical Mechanics} (3rd edition, Addison--Wesley, 2002).

\bibitem[GK10]{GK10}
S.~Grunau, V.~Kagramanova,\emph{Geodesics of electrically and magnetically charged test particles in the Reissner--Nordstr\"om space--time: analytical solutions}, in preparation.

\bibitem[HHLS10]{Hackmannetal10}
E.~Hackmann, B.~Hartmann, C.~L\"ammerzahl, and P.~Sirimachan,
\emph{Test particle motion in the space-time of a Kerr black hole pierced by a cosmic string}, Phys. Rev. {\bf D 82} (2010) 044024.

\bibitem[HL08]{HackmannLaemmerzahl08}
E.~Hackmann and C.~L{\"a}mmerzahl,
\emph{Complete analytic solution of the geodesic equation in Schwarzschild--(anti) de Sitter space--times}, Phys. Rev. Lett. {\bf 100} (2008) 171101.

\bibitem[HL08a]{HackmannLaemmerzahl08a}
E.~Hackmann, and C.~L{\"a}mmerzahl, \emph{ Geodesic equation in Schwarzschild--(anti-)de Sitter space--time: Analytic solutions and applications}, Phys. Rev. {\bf D 78} (2008) 024035.

\bibitem[HKKL08]{Hackmannetal08}
E.~Hackmann, V.~Kagramanova, J.~Kunz and C.~L\"ammerzahl, \emph{Analytic solutions of the geodesic equation in higher dimensions}, {\it Phys. Rev.} {\bf D 78}, 124018 (2008).

\bibitem[HKKL09]{Hackmannetal09}
E.~Hackmann, V.~Kagramanova, J.~Kunz, and C.~L\"ammerzahl, \emph{Analytic solutions of the geodesic equation in axially symmetric space-times}, Europhys. Lett. \textbf{88} (2009) 30008.

\bibitem[HKKL09a]{Hackmannetal010}
E.~Hackmann, C.~L\"ammerzahl, V.~Kagramanova, and J.~Kunz, \emph{Analytical solution of the geodesic equation in Kerr-(anti) de Sitter space-times}, Phys. Rev. \textbf{D 81} (2010) 044020.

\bibitem[HLS10]{HackmannLaemmerzahlSchaffer10}
E.~Hackmann, C.~L\"ammerzahl, and I.~Schaffer, \emph{Analytical solutions of the motion of test particles with spin}, in preparation.

\bibitem[NdSprep]{NdSprep}
E.~Hackmann, V.~Kagramanova, J.~Kunz, and C.~L\"ammerzahl, \emph{Analytic solution of geodesic equations in the Taub--NUT--de Sitter space--times}, in preparation.

\bibitem[MPprep]{MPprep}
E.~Hackmann, V.~Kagramanova, J.~Kunz, and C.~L\"ammerzahl, \emph{Analytic solution of geodesic equations in the Myers-Perry space--time}, in preparation

\bibitem[Hag31]{Hagihara31}
Y.~Hagihara, \emph{Theory of the relativistic trajectories in a gravitational field of Schwarzschild}, Japan. J. Astron. Geophys, {\bf 8} (1931) 67.

\bibitem[Hon07]{Hone07}
A.~N.~W.~Hone, \emph{Sigma function solution of the initial value problem for Somos 5 sequences}, Trans. Amer. Math. Soc. \textbf{359} (2007) 5019 (electronic).

\bibitem[Jor92]{jo92}
J.~Jorgenson, \emph{On directional derivatives of the theta function along its divisor}, Israel J.Math. \textbf{77} (1992), 274.

\bibitem[MP86]{MyersPerry1986}
R.~C.~ Myers, M.~J.~Perry,  \emph{Black holes in higher dimensional space-times}, Ann. Phys. (NY) \textbf{172} (1986) 304.

\bibitem[KKZ10]{KodamaKonoplya2010}
H.~ Kodama, R.~A.~ Konopolya, A.~ Zhidenko, \emph{Gravitational stability of simply rotating Myers--Perry black holes: Tensorial perturbations}, Phys. Rev. D \textbf{81} (2010) 044007.

\bibitem[MP08]{matprev08} S. ~Matsutani, and E. ~Previato,  \emph{Jacobi inversion on strata of the Jacobian of the $C_{rs}$ curve  $y^r = f(x)$}, J. Math. Soc. Jpn. \textbf{60} (2008) 1009-1044; http://www.mittag-leffler.se/preprints/0607/files/IML-0607-41.pdf.

\bibitem[MP10]{matprev10} S. ~Matsutani, and E. ~Previato
 \emph{Jacobi inversion on strata of the Jacobian of the $C_{rs}$ curve
 $y^r = f(x)$ II},
arXiv: 1006.1090v1 [math.AG] 6 Jun. 2010

\bibitem[Mark92]{marku92}
A.~I.~ Markushevich, \emph{Introduction to the classical theory of Abelian functions}
(1992), AMS, Providence.

\bibitem[Mat03]{mat03}
S.~ Matsutani, \emph{Recursion relation of hyperelliptic PSI-functions of genus two},
Int. Transforms Spec. Func. {\bf 14} (2003) 517. 

\bibitem[Nak08a]{nakaya08}
A.~Nakayashiki, \emph{{A}lgebraic expression of sigma functions of $(n,s)$ curves}, arXiv:0803.2083, 2008.

\bibitem[Nor10]{nor10}
T.~Northover,
\emph{Riemann surfaces with symmetry: algorithms and applications}, PhD dissertation, University of Edinburgh, 2010.

\bibitem[Nor10a]{nor10a}
T.~Northover  \emph{Software: Cycle painter}, http://gitorious.org/riemanncycles

\bibitem[{\^O}ni98]{on98}
Y.~{\^O}nishi, \emph{Complex multiplication formulae for hyperelliptic curve of genus three}, Tokyo J. Math. \textbf{21} (1998) 381.

\bibitem[{\^O}ni02]{on02}
Y.~{\^O}nishi, \emph{Determinant expressions for abelian functions in genus two}, Glasgow Math. J. \textbf{44} (2002) 353.

\bibitem[Sag01]{sag01}
B.~E.Sagan, \emph{The symmetric group. Representations, combinatorial algoriths and symmetric fynction}, Second Ed., Graduate texts in Mathematics, \textbf{203} (Springer, New York, 2001).

\bibitem[Tho870]{thomae70}
J.~Thomae, \emph{Beitrag zur {B}estimmung von $\vartheta(0,0,\ldots,0)$ durch die {K}lassenmoduln algebraischer {F}unctionen}, Journ. reine angew. Math. \textbf{71} (1870) 201.

\bibitem[VSP05]{vasupa2005}
M. Vasudevan, K. A. Stevens, D.N. Page  \emph{Separability of the Hamilton--Jacobi and
Klein--Gordon equations in Kerr--de Sitter metrics}, Class. Quantum. Grav. \textbf{22} (2005) 339.

\bibitem[Vanh95]{vanha95}
P.~Vanhaecke, \emph{Stratification of hyperelliptic Jacobians and the Sato Grassmanian}, Acta. Appl. Math. \textbf{40} (1995) 143.

\bibitem[Wei893]{weierstrass893}
K.~Weierstra{\ss}, \emph{Formeln und {L}ehrs{\"a}tze zum {G}ebrauche der elliptischen {F}unctionen}, bearbeitet und herausgegeben von H. A. Schwarz (Springer, 1893).

\bibitem[W09]{Will09}
C. ~Will, \emph{Testing the general relativistic ``no-hair'' theorems using the galactic center black hole ${\rm SgrA}*$}, Astrophys. J. Lett. \textbf{674} (2008) L25.

\end{thebibliography}
\end{document}